\documentclass[3p, onecolumn]{elsarticle}
\journal{Physics of Life Reviews}
\usepackage{amssymb}
\usepackage{graphicx}
\usepackage{amsmath}
\usepackage{enumerate}
\usepackage[usenames,dvipsnames]{color}

\biboptions{sort&compress}

\title{Evolutionary game theory and the tower of Babel of cooperation: Altruism, free-riding, parasitism and the structure of the interactions in a world with finite resources.}

\author[UAB]{Rub\'en J. Requejo-Mart\'inez\corref{cor1}}
\ead{rubenfisico@yahoo.es}
\address[UAB]{Department de F\'{\i}sica, \\
Universitat Aut\'{o}noma de Barcelona, \\ 08193 Bellaterra, Barcelona, Spain}
\cortext[cor1]{Corresponding author}

\begin{document}

\begin{frontmatter}

\begin{abstract}

The study of the evolution of cooperative behaviours --which provide benefits to others-- and altruism --which provides benefits to others at a cost to oneself-- has been on the core of the evolutionary game theoretical framework since its foundation. The fast development of the theory during the last years has improved our knowledge of the issue, but carried attached a diversification of concepts which affected communication between scientists. Furthermore, the main root of conflict in the struggle for life identified by Darwin, the limited amount of resources present in any ecosystem, which is assumed to keep a constant population size in most game theoretical studies, has only recently been taken into account as explicitly influencing the evolutionary process. This review concerns about both issues, the conceptual diversification during the last years and the new results of the resource dependent models. In extenso: After a historical introduction, a review of the most important concepts is carried out. Then it is shown that pairwise interactions and additive fitness determine prisoner's dilemmas (PDs) or harmony games, that two altruists interacting together may determine a PD, and that the interaction environment of the most cooperative and less selfish individual in any population is always a PD. After that, it is shown that in addition to altruists versus free-riders, the combination of free-riders and parasites determines a fundamentally different PD. Computer simulations are then carried out to show that random exploration of parasitism, free-riding and altruism enables coexistence of the three strategies without the need of reciprocating, punishing or rewarding strategies. To finish, the problem of the limitation of resources is reviewed, showing that trade-offs between rate and yield in resources use do not allow for coexistence of cooperative and defective strategies in well-mixed situations, but that a certain feedback effect between resources use and parasitic benefits may allow for it, which represents an exception --the first one reported to my knowledge-- to the competitive exclusion principle.

\end{abstract}


\end{frontmatter}

\tableofcontents

\section{Introduction}
\subsection{Evolution and natural selection}
The publication in 1859 of the book \emph{On the Origin of Species by Means of Natural Selection, or the Preservation of Favoured Races in the Struggle for Life} by Charles Darwin \cite{darwinl:1859} marked the end of an old era dominated by religious beliefs in which the human being --specially the occidental human being-- was in the middle of the universe of creation. The idea of evolution, independently found by Wallace and Darwin himself at a time in which scientific circles questioned creationist theories, laid the foundations for scientific proof of the revolutionary ideas: man, as well as any other living organism, is the product of a long evolutionary process involving small changes and selection.

The first ideas related to evolutionary processes root on the late 1700's and early 1800's lively debate on the formation and shaping of the earth, where some geologists, as James Hutton and Charles Lyell \cite{hutton:1788, hutton:1794, hutton:1795, lyell:1853}, proposed that very slow long term processes of microscopic change driven by natural forces, as wind and water flow frictions or temperature changes, where the actual cause for the observed earth structure (Hutton \cite{hutton:1794} and Lyell \cite{lyell:1853}, specially the first, already suggested the applicability of such ideas to the study of biological processes). Johann W. Goethe \cite{goethe:1790} also noticed that, given the morphological similarities between all plants, they might have developed by metamorphosis from an equal original form or ur-plant. In this context, Jean-Baptiste Lamarck made the first proposal of a biological evolutionary process giving rise to new species \cite{lamarck:1809}.

At the same time as the evolutionary geology debate was held, economist T. Malthus ideas on population growth, overuse and competition for resources spread, reaching Wallace and Darwin. Malthus argued that any increase in available resources in society would lead to a subsequent increase in the population, until the same original subsistence per-capita amount of resources was reached \cite{malthus:1803}. Darwin and Wallace thought that, if this was to happen in human societies in which individuals may restrain their own reproductive and consumption rates, it would happen still more intensely in nature, where animals were thought not to do so, neither to have the ability to increase their resource supply.

The application of geological and economic born ideas, together with the competition triggered by the finiteness of resources, led Darwin and Wallace to the conception of the evolutionary process in which individuals reproduce and give birth to similar --but not equal-- offspring, and natural selection allows for the survival and spread of the best adapted traits, those associated to the fittest individuals, understood as the most successful from a reproductive and survival perspective.

However, Darwin himself realised the paradox implicit in natural selection acting at the individual level: any living being exploiting others would have a net evolutionary advantage over those individuals which assume some reproductive cost in order to produce a benefit on the rest; thus, the evolution of cooperative and altruistic behaviours seems to be doomed, which led to the imposition of the most dramatic view of the struggle for life in some scientific circles \cite{huxley:1888} in the late 1800's and early 1900's, and to the denial of the existence of cooperative benefits from an evolutionary perspective.

The evolution of cooperative behaviours was analysed from an opposed point of view in 1902 by Piotr Kropotkin in his book \emph{Mutual Aid: A Factor in Evolution} \cite{kropotkin:1902}, where he gathered a collection of articles in which he showed that cooperation, present both in animals and humans, is an important factor to take into account from an evolutionary perspective. He did not neglect natural selection, but argued that the struggle against an inclement nature favoured the evolution of mutual aid instead of fight between con-specifics.

Despite some experimental results showing the benefits of group formation against under-crowding in unfavourable environments \cite{allee:1927b}, the debate continued, with supporters on both sides. It would still take another seventy years for theoreticians to develop a theory consistent with the observed results, and able to account for the evolution of cooperation.

 	\subsection{Genes, populations, relatedness and assortment}

At the time Wallace and Darwin proposed the theory of natural selection it was still unknown how heredity of traits between parents and offspring was realised. The works on heredity of Mendel, published in 1866, remained unknown until the beginning of the 1900's, when three European scientists --Hugo de Vries, Carl Correns, and Erich von Tschermak-- found similar results and rediscovered it. They found experimental results proving the transmission of discrete traits between parents and offspring, and a few years later the term gene was coined.

Even when the discovery of the DNA as carrier of the genetic information should still wait until the 1940's, the knowledge of the existence of such inherited information as discrete traits allowed theoreticians --S.Wright, R. A. Fisher and J.B.S. Haldane \cite{fisher:1930,haldane:1932,wright:1968}-- during the 1920's and 1930's to develop a mathematical framework for the evolution of gene frequencies and their associated traits within populations, in what is now called population genetics.

In this context, the importance of relatedness as a measurement of genetic similarity came into attention as a possible factor allowing for the evolution of altruistic sacrifice, as illustrated by J.B.S. Haldane statement that he would not risk his life for saving a drowning brother, but he would do it for two brothers or eight cousins. This intuitive statement was formalised mathematically in 1964, when W.D.Hamilton published his seminal works on kin selection \cite{hamilton:1964a, hamilton:1964b}, founding the inclusive fitness theory.

Inclusive fitness theory is based on the assumption that, although selection is carried out at an individual level, it is the genes which are actually selected, and equal genes are indistinguishable from an evolutionary perspective. Thus, the fitness of a behavioural trait, which is a measurement of its reproductive value and directly related to the number of offspring it will produce, is not only that of the trait in the individual, but also the addition of the effects of its behaviour on the fitness of all other individuals carrying such trait. Furthermore, it is usually assumed that evolution acts so as to maximise the inclusive fitness of the individuals.

With the previous assumptions applied to identical genes by descent, i.e. genes which are a perfect replica of those of a common ancestor, Hamilton proved that altruistic traits which imply a cost $-c<0$ to the actor might evolve whenever the benefit $b>0$ of the altruistic behaviour is directed towards individuals whose relatedness $r$ --a measurement of genetic similarity-- fulfils the so called Hamilton rule
\begin{equation}
\label{eq:hamilton}
r>\frac{c}{b},
\end{equation}
situation in which the inclusive fitness effects of the costly action to the altruistic donor trait are outweighed by the benefits accrued on similar-enough individuals (see \cite{frank:1997} for a detailed explanation of the meaning of $r$).

This results were expanded in 1970 by G.R. Price \cite{price:1970}. He proved that, if the fitness of the individuals carrying a trait $i$ of value $z_i$ at time $t$ is given by $w_i=\bar{w} q_i'/q_i$, where $q_i, q_i'$ are the frequencies of individuals carrying such trait at times $t$ and $t'=t+1$, and $\bar{w}$ is the mean population fitness, then the variation of the mean value of such trait $\Delta \bar{z} = \bar{z}'-\bar{z}$ fulfils
\begin{equation}
\label{eq:price}
\bar{w} \Delta \bar{z} = Cov(w_i, z_i) + E(w_i \Delta z_i),
\end{equation}
where $Cov(w_i, z_i)=E(w_i z_i)-E(w_i) E(z_i)$ is the covariance between fitness and trait value, and $E(X)$ is the expected value of $X$.

The Price equation applied to the evolution of altruistic behaviours results in the Hamilton's rule, but its interpretation changes. The fundamental feature which allows for the evolution of altruism is no longer genetic relatedness, but the assortment between altruistic behaviours \cite{fletcher:2009}, i.e. the fact that enough benefits given by altruists are enjoyed by other altruists. It is interesting to note that the Price equation is derived from simple mathematical arguments, for which reason it is always a valid equation in any system in which standard mathematics apply, but as some authors argue, this generality deriving from its mathematical truth makes this equation content empty: It does not say anything else about the real world which is not already implicit in the structure of mathematical statistics.

		\subsection{The mathematics of games, or how to model behaviours}

The origin of game theory dates back to the 1920's, when John von Neumann published a series of articles on the issue, and the latter book \emph{The Theory of Games and Economic behaviour} in collaboration with Oskar Morgenstern. Its development continued during the rest of the century, attracting initially  the attention of economists and politicians. 

Game theory focuses on the study of cooperation and conflict between rational decision makers interacting together, i.e. individuals who possess information about the possible outcomes of the interactions (the game) and decide how to act according to it. The specification of a behaviour of an individual in any situation is called strategy, and the outcome of every interaction depends on the strategies chosen by all interacting individuals (players). The main goal of game theory is to predict which strategies will be played by each player and the associated distribution of benefits, for which reason John Forbes Nash introduced in 1951 the concept later called Nash equilibrium. A Nash equilibrium happens whenever none of the players increases its benefits by changing its actual strategy. In this way, if one assumes that individuals are rational, the Nash equilibrium represents the outcome of the interaction.

Full rationality was initially assumed in game theoretical models. This means that individuals are rational, and take into account that their interacting partners are rational as well. This assumption is often unrealistic, on the one side because information might not be fully available or costly to acquire, and on the other side because it leads to an infinite iteration of the form I know that you know that I know that you know..., for which reason the concept of bounded rationality was introduced into economic game theory. 

Bounded rational individuals do not possess any longer all the information of the system, or cannot process it, and their behaviour is influenced only by a few variables related to the situation. This point of view seems more appropriate to describe the real world, in which animals (including humans) neither have infinite perception of the reality surrounding them, nor --often-- the time to process all important information before interacting. And this is specially important in a situation in which life, reproduction and death come into play.

All previous concerns illustrate the necessity of a dynamic framework in which to embed the game theoretical analysis (this was already suggested by Nash in his doctoral thesis), which led to the birth of evolutionary game theory.

		\subsection{Evolutionary game theory versus inclusive fitness theory: modelling behavioural evolution.}

In 1973 J. Maynard-Smith and G.R. Price published a paper in which they reinterpreted the payoffs of the game as fitness changes of the individuals, and thus in their reproductive capacity. The static view of game theory was transformed into a dynamic framework, and the Nash equilibrium concept was changed for that of evolutionary stability, to refer to those populations which, once established, cannot be invaded by just a few mutant individuals \cite{maynard-smith:1973, maynard-smith:1982}. In this way, the branch of biology called evolutionary game theory was born.

In the following years the replicator equation, 
\begin{equation}
\label{eq:replicator0}
\frac{dx_i}{dt} = x_i (f_i - \bar{f}),
\end{equation}
became the main tool for analysing the dynamics resulting from the evolutionary processes \cite{taylor:1978, schuster:1983}. In this equation $x_i$ is the fraction of individuals following strategy $i$, $f_i$ is their fitness and $\bar{f}$ the mean population fitness. This equation describes the frequency-dependent dynamics of infinite --or very big-- populations of replicating individuals, i.e. the dynamics in a fitness landscape which depends on the population composition. On its mutation regarding version, the so called replicator-mutator equation has been proposed as a dynamical equation describing the entire evolutionary process of reproduction, mutation and selection. 

The replicator equation is easily connected with ecological models via the Lotka-Volterra equations \cite{hofbauer:1998}. In particular, it has been proven that there is a transformation between the variables of the Lotka-Volterra equation for $n-1$ species and the replicator equation for $n$ phenotypes which results in the same orbits \cite{hofbauer:1998}. The replicator equation has also been proven to be equivalent to the Price equation if one assumes that the traits involved in the latter are indicator functions which take value $1$ for equality of two indices and $0$ otherwise, and that the average of a trait remains constant in the absence of selection \cite{traulsen:2010}.

The debate on the most fundamental approach to formulate mathematical models of evolution (evolutionary game theory versus inclusive fitness theory) has returned to be a hot topic during the last years \cite{nowak:2010,vanveelen:2010,nonacs:2010,krakauer:2010,doebeli:2010,abbot:2011,strassmann:2011, ferriere:2011,herre:2011,boomsma:2011,bourke:2011,nowak:2011b,fromhage:2011,damore:2012,alger:2012, simon:2012,vanveelen:2012}. Although recent studies argue or show that both disciplines provide the same results in some situations \cite{lehmann:2006b,marshall:2011}, this has not been proven in a broader scope, and it has been recently suggested that they are indeed fundamentally different \cite{vanveelen:2012,traulsen:2010}. One of the main differences is that the Price equation is not always dynamically sufficient, i.e. cannot be used to predict all variables in future temporal steps knowing the previous state of the system, and thus to calculate trajectories in the phase space, which suggests that the evolutionary game theoretical framework might be more appropriate whenever we want to describe the dynamics of the system in detail and calculate associated quantities, as fixation probabilities or invasion times \cite{traulsen:2010}. Furthermore, the evolutionary game theoretical framework intrinsically concerns about frequency dependent selection \cite{nowak:2006a}, i.e. situations where the evolutionary process depends on the strategy frequencies present in the population, and allows for a general treatment for any intensity of selection \cite{nowak:2004a}. 

It is worth to remark the difference between the assumption of weak selection in both frameworks: While this assumption in game theory refers to the ``importance'' (influence) of the game (interaction structure) in the reproductive capacity of the individuals, in inclusive fitness it usually refers to a very small difference between the behaviours involved in the evolutionary process; the latter assumption usually leads to the neglection of non-linear terms and induces an underlying mathematical structure which have been argued to constrain the ability to represent mathematically the evolutionary process \cite{wild:2007,traulsen:2010}. While this may be the right representation for most genetically related evolutionary systems --which assume that mutations lead mostly to small phenotypic changes--, it may fail for social evolution (see \cite{traulsen:2010} for a detailed discussion).

In the last two decades the evolutionary game theoretical framework has been expanded to cover a much broader range of situations. In addition to the study of evolution in nature, culture and society have been regarded from an evolutionary perspective. In cultural evolution and social learning contexts, strategies are no longer linked to genes, and the reproduction and death process are reinterpreted as a change of strategy during the lifetime of the individual. Furthermore, new microscopic update rules have been introduced to represent specific learning methods and reproduction-death processes, broadening the range of applications to include sociological and psychological research. However, with the development and expansion of the game theoretical framework, evolution has acted upon some of the key concepts that initially originated it. In this review, I first focus on these key concepts, trying to unify different nowadays perspectives, and using them to illustrate some interesting situations; then I review and summarize present results that go back to the original problem pointed out by Malthus and Darwin as one of the origins of competition in nature: the limited amount of resources present in our world. 

\section{Definitions and methods}
\subsection{Relative definitions of cooperative and selfish behaviours, a general framework for cooperation: Disentangling selfishness and cooperativeness.}\label{sec:reldef}

The different definitions of cooperation and altruism used by scientists have sometimes led to misunderstandings, affecting communication between them and knowledge transfer. A recent attempt to start a debate on the use of such concepts seems to have left things unchanged \cite{west:2007,bergmuller:2007,taborsky:2007}, with an unclear definition of the concepts, or even worst, with a multiplicity of definitions which use the same words to refer to different things. In what follows some simple definitions are given, which relate to classical definitions and clarify some common misunderstandings between scientists, as those deriving from the mix of group versus individual behaviour, of the use of different time frameworks or of the assumption that selfishness and cooperativeness are correlated (some of the misunderstandings are analysed in section~\ref{sec:misund}).

The concept of cooperation, rather than a fixed concept that allows for a clear definition of what cooperative individuals are, is a contextual dependent concept, as strategies can be defined as cooperative or not only in relation to other strategies involved in the evolutionary process \cite{bull:1991,dugatkin:1997}  (this is expanded with examples in section~\ref{sec:PD}). This suggests that, instead of trying to define strategies as cooperative or defective, a gradual definition for cooperation might be established according to the following two principles: 

\emph{
\begin{itemize}
\label{def:coopbeh}
\item A strategy A is more cooperative than another one B whenever it is more beneficial to interact with A than with B from a receiver's point of view.
\item A cooperative act does not reduce the fitness of the recipient of the act.
\end{itemize} 
}
The first principle to define a cooperative behaviour or act is purely relative, as it is based on the perception of benefits of one individual when interacting with others, while the second one may have a relative or an absolute character depending on the choice of the reference system on an absolute or a relative basis (the discussion of this point is expanded in section~\ref{sec:nullpoint}). In this way, if two individuals which are regarded as cooperative according to these two principles interact together, the interaction is a cooperative act. Furthermore, the principles just given to define an act as cooperative are completely consistent with the general definition of cooperation, which might be stated as follows:
\emph{
\begin{itemize}
\item Cooperation is the non-forced action or process of working together for a common purpose or benefit.
\end{itemize} 
}
Remember that the referred benefits in a biological or social context are usually quantified as fitness, i.e. as a measure of the reproductive capacity of the individual or the behavioural trait.

Let me remark that in order to define a behaviour as cooperative according to the previous principles one has to look at the effect of such behaviour on the co-player, and not on oneself: Self-benefits are important for the definition of selfishness, not of cooperativeness. In analogy with the concept of selfishness, a relative definition for the concept of selfish behaviour may be stated as:

\emph{
\begin{itemize}
\label{def:selfishbeh}
\item A strategy A is more selfish than another one B whenever it is more beneficial to act following strategy A than B from an actor's point of view.
\item A selfish act increases the fitness of the the actor.
\end{itemize} 
}
And the general definition of selfishness as:
\emph{
\begin{itemize}
\item Selfishness is the quality of individuals which concern excessively or exclusively with oneself, seeking or concentrating on one's own advantage, pleasure, or well-being without regard for others.
\end{itemize} 
}

As before, the first principle to define a behaviour as selfish is purely relative, based on the perception of benefits of the actor following two different strategies, while the second one may have a relative or an absolute character depending on the choice of the origin of the reference system. Now, complete classification of strategies can be done following the definitions of cooperative and selfish behaviours (Fig.~\ref{fig:diagramanp}). Note that this classification includes cooperative behaviours which are also (at least to some extent) selfish, as the cooperator gets a self-benefit for cooperating; this behaviours, which are usually called mutually beneficial or mutualistic, include weak altruism, which might be regarded as selfish cooperation, but in which the net benefit of the cooperator is less than the net benefit of the co-player, i.e. in which the individual is more cooperative than selfish (see \cite{wilson:1990} for a nice review of the concepts of weak and strong altruism in groups. In the following altruism refers to strong altruism). 

In summary, \emph{cooperativeness} refers to fitness variations in co-player, requiring for absolute cooperation such effects to be non-negative when compared to the neutral case, i.e. the null point of the reference system (Fig.~\ref{fig:diagramanp}). In the same way, \emph{relative selfishness} refers to fitness variations in oneself, with the requirement for absolute selfishness of such effects to be positive when compared to the null point of the reference system. In this way, cooperativeness and selfishness are different features of a behaviour, not necessarily dependent on each other. Note however that absolute cooperativeness and selfishness are not absolute in the sense of applicable for any framework, but within the selected framework, which in most cases is a relative framework which uses an individual behaviour, or the mean population fitness, as origin of the reference system. The discussion on the choice of the framework is expanded in section~\ref{sec:nullpoint}.

	\subsubsection{Remarks on the definitions and common misunderstandings}
	\label{sec:misund}
The second requirement for a cooperative behaviour states that the benefits of cooperation must be at least the absence of fitness losses due to aggressions, and from this ground, any positive benefit, as those created by altruists; a slightly more stringent definition requires positive effects (even if infinitesimal) on the partner \cite{bergmuller:2007}. However, the definitions given above account for peace as a social good \cite{rankin:2007}, and thus conflict-avoiding, pacific and pacifist behaviours as cooperative. Note that, although when compared with altruism, free-riding strategies (neutral individuals) are regarded as defective, as they receive the altruistic benefits at no cost, the cost paid by altruists is not a consequence of the free-riding strategy, but of the altruistic behaviour itself; furthermore, as I will show in section \ref{sec:coexistence}, in a context in which there exist aggressive, parasitic strategies, the existence of free-riders and altruists allows for their survival, preventing the sure extinction of any of those strategies when interacting with parasites in well-mixed populations, i.e. in the mean field limit in which every individual interacts with any other with equal probability. 

Note also that the previous definitions may apply to cooperation in the short-term, i.e. interaction after interaction, as well as in the medium term (the result of a certain number of interactions) and in the long term (lifetime consequences of the behaviour \cite{west:2007}), but that such time frame should always be specified \cite{brosnan:2010, jensen:2010}, as strategies which are regarded as non-selfish and cooperative (altruistic) in the short term may turn to be selfish in the long term. As an example of the misunderstandings that may arise due to the lack of specification of the temporal framework, let us analyse the discussion on the the so called reciprocal altruism \cite{trivers:1971}.

Altruism is a behaviour which incurs costs to the actor in order to produce benefits in the recipient of the act, being a purely cooperative and non-selfish behaviour. However, according to the definitions above, some acts that might be seen as altruistic in the short term may turn to be mutually beneficial in the long term, as it happens for behaviours that reciprocate altruistic acts \cite{trivers:1971, axelrod:1981}: they might be called mutually beneficial (selfishly cooperative) in the long term, but altruistic (non-selfish and cooperative) in the short, whenever individuals have no information and cannot predict the outcome of the next interaction. Thus, the discussion on whether reciprocal altruism is actual altruism or not was triggered by a choice of different time frames, not of different concepts: those who see it as altruism were looking to the short term, while those considering it a selfish strategy where looking to the lifetime consequences. Note however that independently of the time frame chosen, reciprocal altruistic behaviours are always regarded as cooperative according to the definitions, being the difference their degree of selfishness, which changes from negative (non-selfishness) to positive with increasing time.

The misunderstanding in the reciprocal altruism debate roots on the mix of two different points of view: the short term vs. the lifetime consequences. This relates directly with two of the major problems in biology, as expressed by N. Tinbergen in 1963 \cite{tinbergen:1963}: the causation and the survival value problems (see also \cite{mayr:1961} and the discussion in \cite{west:2007}). The causation problem might be colloquially said to relate to proximate explanations, or to answer ``how'' questions, while the survival value may be related with ``why'' questions, or ultimate explanations. In this way, the altruistic reciprocating behaviour is the answer to how an animal may behave in order to promote cooperation in the short term. Why such behaviour evolves is because it produces enough assortment so as to be promoted by evolution on the long term, but this can only be stated after the entire life history of the altruistic behavioural trait is taken into account. 

Other mistake found in the literature comes from assuming that altruism and selfishness are different degrees of the same function. In this way, parasitic behaviours have been called selfish behaviours \cite{west:2007}. It is true that parasitism is the paradigmatic example of a selfish non-cooperative behaviour. However, as it has been shown before, selfishness and cooperativeness are not mutually exclusive (Fig.\ref{fig:diagramanp}) and it is perfectly possible to have selfish strategies that produce some benefit in the co-player, and are thus cooperative. For this reason, identifying selfishness with parasitism restricts the broader scope of what selfishness is and should be avoided.

Further mistakes come from the confusion of the concept of cooperation at the individual level, i.e. cooperative behaviour, and cooperation as a collective action --the broader definition. This confusion, which was already pointed out by Dugatkin in 1997 \cite{dugatkin:1997,bergmuller:2007} is however not cleared in some later papers \cite{west:2007}. 

Some authors have also tried to relate the definitions of altruism and cooperation with the evolutionary consequences of such behaviours \cite{west:2007,brosnan:2010}, and more specifically, by the way they are selected for. However, such definitions fail whenever a behaviour is counter selected, and may induce further confusion allowing for situations in which altruism and cooperation seem exclusive from each other (see appendix~\ref{sec:definitions} for a short summary of definitions in the literature).

To finish, let us note that, as pointed out by some authors \cite{brosnan:2002}, even if intentionally directed towards cooperation, cooperative behaviours may fail to produce the expected benefits. In this case, the long term behaviour, influenced by the probability of success, can be used to define it as cooperative or not. Note that with this extension of the definition, acts that cannot reduce other individuals fitness are always regarded as cooperative, even if they always fail to provide the expected benefits.

Further discussion about the intentionality and non-intentionality of the cooperative behaviours and its relationship with the emergence of ecosystems can be found in appendinx~\ref{app:remarks}. 

	\subsubsection{Discussion on the null point of the reference system}
	\label{sec:nullpoint}
Along the text the prisoner's dilemma (PD, explained in the next section) will be extensively used to illustrate the cases of cooperation by restraining from conflict (free-riders versus parasitism), and that of social goods formed by cooperation (altruism versus free-riding). However, it is good to remark that both situations, which are comprised in the definition of cooperation provided at the beginning, are relative situations, as they refer to fitness changes in reference to a certain behaviour which determines the null point. In the PD examples along the text, the reference used will be a passive, neutral individual, called free-rider, as it produces no fitness variations on others, nor on itself, but receives the actions of the co-players. 

\begin{figure}
\begin{center}
\includegraphics[width=100mm]{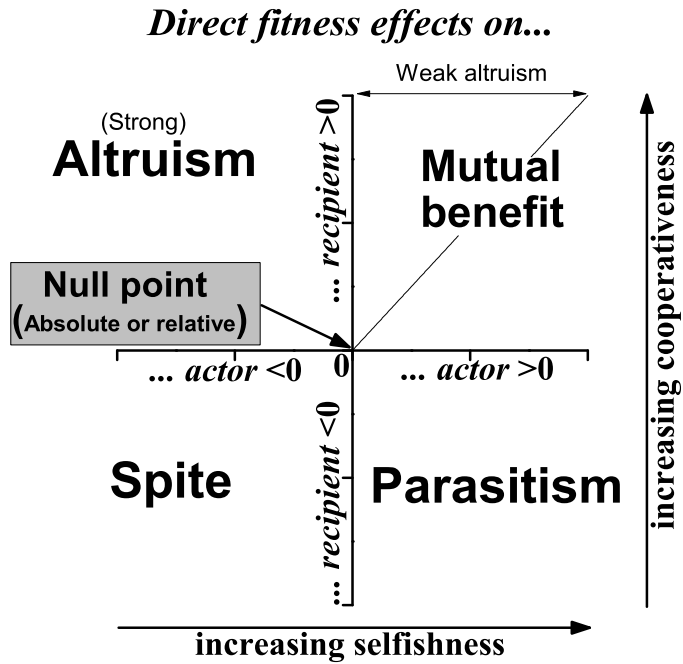}
\caption{Classification of strategies in the case of direct interactions as a function of the direct fitness effect of the action on the actor itself and on the recipient. The arrow on the right shows the increase in cooperativeness; the increase in selfishness is shown by the lower arrow. According to the definition, only non-damaging strategies (upper quadrants including the x-axis) should be called cooperative, while only strategies increasing their own fitness should be called selfish. Neutral individuals (free-riders) determine the null point, altruism is non-selfish cooperation, mutualism is selfish cooperation, spite is non-selfish and damaging, and parasites are selfish non-cooperative individuals. The combination of any two of these strategies determines either a Prisoner's Dilemma (PD) or a Harmony Game (HG), as shown in section~\ref{sec:onlyPDandHG}. \label{fig:diagramanp}}
\end{center}
\end{figure}

Whenever the null point is determined by a passive behaviour, it can be easily used as null point or origin of the reference system, as the one in figure~\ref{fig:diagramanp}. However, it might not be always easy (maybe not even possible) to find such reference points in other situations, specially whenever all individuals are active, e.g. situations in which all individuals are reproducing. In this case individuals refraining from reproduction reduce their own reproductive fitness and increase that of others, being thus altruists, while those that increase their reproductive rate and decrease the fitness of others would be parasites. However, as in this case all individuals are actively reproducing, it may happen that no fixed baseline fitness can be measured as a neutral reference. In such cases two possibilities appear: first, choosing a constant arbitrary reference for the origin of the behavioural reference system; second, using the mean population fitness (or other variable) as reference. The latter case seems to be a very opportune choice, as the replicator equation (analysed in Sec.~\ref{sec:DynamicalOutcomes}), which is usually assumed to describe the system's dynamics in evolutionary game theoretical studies, depends on the difference of the individual and the mean population fitness. In this special case the reference system (Fig.~\ref{fig:diagramanp}) would have a time dependent origin, and thus cooperativeness and selfishness degrees would not only be relative, but also time dependent.

	\subsection{Non-cooperative symmetric games}
	\label{sec:nonCgames}
In evolutionary game theory the essence of the interactions between individuals is captured in a payoff matrix, where each entry represents the outcome of the interaction between a focal individual behavioural type (rows) and the interaction partner behavioural type (columns) (this refers to normal form games; see \cite{cressman:2003} for extensive game respresentations). A general payoff matrix for non-cooperative symmetric games between two strategies, A and B, may be written as 
\begin{equation}
\label{eq:PayoffMatrixGen}
\begin{tabular}{l|cc}
 & A & B \\
\hline
A & $R$ & $S$ \\
B & $T$ & $P$
\end{tabular}
\end{equation}
and depending on the parameters, four different kinds of non-cooperative games are defined, where the cooperative behaviour is only promoted whenever it is the most selfish behaviour (this will be explained in deep later). If we assume that behaviour A is the cooperative behaviour, the four possible games are the prisoner's dilemma (PD), the snowdrift (SD) or chicken game, the stag hunt (SH) game and the harmony (HG) game.

The \emph{prisoner's dilemma} is named after the dilemma in which two prisoners are asked to incriminate the other as participant in a robbery. If both individuals incriminate each other, each one spends 3 years in prison. If only one of them incriminates the other, they spend respectively 0 and 5 years in prison. If none of them denounces the other, both spend 1 year in prison. Thus, it is better for them if none incriminates the other (both cooperate), and both get free after 1 year, than if both testify against the other (both defect) and spend 3 years in jail. However, rational behaviour leads to mutual defection, as from a self-maximising point of view it is always better to incriminate: if the other one does not, incriminating is worth freedom instead of 1 year in prison; if the other one incriminates you, incriminating reduces in two years the time in prison, from 5 to 3 years. The general payoff ordering in Eq.~\ref{eq:PayoffMatrixGen} for a PD is $T>R>P>S$ (note that, as spending time in prison is bad, the payoffs should be taken as negative, i.e. 0, -1, -3, -5).

The \emph{snowdrift} or chicken game is named after two different situations which result in the same payoff ordering: $T>R>S>P$. The snowdrift refers to a situation in which two drivers are stuck in the middle of a snowdrift in the morning. They can both cooperate to shovel the snow (thus sharing the working efforts) and get home for lunch. If both decide to defect, however, they will have to wait until the afternoon for the sun to melt the snow, and they will arrive home for dinner. If just one of them decides to shovel the snow, they will arrive home after lunch, but the individual cooperating is the only one paying the cost of working. The chicken game refers to a situation in which two drivers drive towards each other, and the one that lasts longer without changing direction wins a prize. If both of them change direction, the prize is shared and none gets hurt. If none of them changes direction, they crash, break their cars, and get injured. The structure of the payoff ordering in this case determines that the best option is to do the opposite of the co-player: If the other one does not cooperate, you should, thus avoiding getting stuck in the snow until dinner, or involved in a car accident; if the co-player cooperates, it is better not to cooperate, in order to save the effort of shovelling the snow, or to win the entire prize. 

The \emph{stag hunt} refers to a situation in which two individuals go hunting together. They can choose between cooperating to hunt a stag, which requires coordinated work and long time waiting, or going each one on their own to hunt rabbits, which is a less worthy prize, but easier to do. The SH game is a coordination game, as the best option is doing the same as the co-player. If both cooperate, they get the stag and the highest possible reward, while if they go hunting for rabbits, they both get a lower reward. However, if one of them goes hunting a stag and the other one goes hunting for rabbits, the first one will not get the stag, while the second one will hunt more rabbits as he will keep all of them instead of sharing with the co-player. The payoff ordering determined by the SH situation is $R>T>P>S$.

The \emph{harmony} game represents a situation in which cooperating is the best outcome, as it maximizes instant payoffs from an individual point of view. In this case the payoff ordering is $R>T>S>P$, and thus, irrespective of the co-players strategy, cooperating increases ones own payoff (from $P$ to $S$ if the co-player defects, from $T$ to $R$ if the co-player cooperates). Thus, this situation does not represent a dilemma for the evolution of cooperation.

	\subsection{Evolutionary dynamics}

	\subsubsection{Invariances of the dynamics}
	\label{sec:Invariances}
If the evolutionary dynamics depend on payoff differences between strategies, or between strategies and mean population payoff, as for the replicator equation~\eqref{eq:replicator0}, or other updating rules \cite{szabo:1998,traulsen:2006b,requejo:2012c}, the dynamics is invariant under the addition of a constant to any column in the payoff matrix. Let us prove it for the general case of $N$ different strategies.

For payoff differences between two strategies, writing $M_{ij}$ for the element in row $i$, column $j$ of the payoff matrix, i.e. the payoff of strategy $i$ interacting with $j$)
\begin{equation}
\pi_i - \pi_j = \sum_{k=1}^{N} x_k M_{ik} - \sum_{k=1}^{N} x_k M_{jk} 
= \sum_{k=1}^{N} x_k (M_{ik} - M_{jk})
\end{equation}
and as the difference in the last term is for payoffs in the same column, i.e. column $k$, the addition of any constant to the entire column leaves the result invariant.

For dynamics ruled by the difference of strategies payoff and the mean population payoff,
\begin{equation}
\pi_i - \overline\pi = \sum_{k=1}^{N} x_k M_{ik} - \sum_{j=1}^{N} x_j (\sum_{k=1}^{N} x_k M_{jk}) 
\end{equation}
then, if one adds a constant $t$ to all payoffs in column $m$, one gets an extra term $x_m t$ from the first summation, and $-\sum_{j=1}^{N} x_j (x_m t) = - x_m t$ from the second one; both vanish as they have different sign.

	\subsubsection{Dynamical outcomes of the replicator dynamics for two-strategy symmetric games}
\label{sec:DynamicalOutcomes}

The prisoner's dilemma (PD) is one of the four possible two-player, two-strategy symmetric games. In the PD the payoff ordering is $T>R>P>S$ (see equation~\eqref{eq:PDmatrix}). As it has been discussed, this makes defection more beneficial from a self-maximising point of view in any interaction. The other three games are the snowdrift (SD) game, corresponding to $T>R>S>P$, the stag hunt (SH) game, corresponding to $R>T>P>S$, and the harmony game (HG), where $R>T>S>P$. In the following, I make use of the replicator equation~\eqref{eq:replicator0} to show the dynamical outcomes of the four games, and how the PD represents the most difficult case for the evolution of altruism in large, well-mixed populations (mean field limit in which every individual interacts with any other with equal probability).
 
The replicator equation~\eqref{eq:replicator0} depends on the difference between the mean payoff of the strategy and the mean population payoff. Therefore, as shown in section~\ref{sec:Invariances}, the addition of a constant to any column of the payoff matrix leaves the dynamics invariant. The payoff matrix for two strategies may thus be written (subtracting the diagonal term in each column) as
\begin{equation}
\label{eq:PayoffMatrixSimp}
\begin{tabular}{l|cc}
 & A & B \\
\hline
A & $0$ & $a$ \\
B & $b$ & $0$
\end{tabular}
\end{equation}
where $a=S-P$ and $b=T-R$ in the terminology of equation~\eqref{eq:PayoffMatrixGen}.

Since the fractions of strategies add up to one, $x_a + x_b = 1$, the entire dynamics may be expressed with just one differential equation, let us say that for $x \equiv x_a$:
\begin{equation}
\label{eq:replicator2strat}
\frac{dx}{dt} = \dot{x} = x(1-x)[a-(a+b)x]
\end{equation}
In this case factor $a$ corresponds to the per-capita growth rate of strategy $A$ when rare, i.e. 
\begin{equation}
a=\lim_{x \rightarrow 0} \frac{\dot{x}}{x} = \left. \frac{d\dot{x}}{dx} \right|_{x=0}
\end{equation}
It can be proven in a similar way that $b$ is the per-capita growth of strategy $B$ when rare in the population. These two factors are important as we may assure if a strategy is evolutionarily stable, i.e. resistant to invasion by a small fraction of mutants, just by looking to them; if the per-capita growth rate of the invader strategy is not positive, it will not increase its presence in the population.

Generally, there will be three solutions for equation~\eqref{eq:replicator2strat} with parameters $a,b\neq 0$. Two of them represent always valid solutions, i.e. $x=0$ and $x=1$. The third one
\begin{equation}
\label{eq:repeqsol}
x^* = \frac{a}{a+b}
\end{equation}
represents a valid solution only when it takes values $x^*\in [0,1]$, and it may be a stable or unstable fixed point.

The outcomes of the dynamics are summarised as follows:
\begin{enumerate}[i.]
\item \emph{Neutral stability}. If $a=b=0$ every point in $[0,1]$ is a rest point. In this case there is no evolution and the fraction of individuals $x$ remains constant in time.

\begin{center}
\begin{minipage}[b]{0.5\linewidth}
\begin{picture}(160,10)(0,0)
\thicklines
\put(30,2){\line(1,0){80}}
\multiput(30,2)(4,0){21}{\circle*{3}}
\end{picture}
\end{minipage}
\end{center}

\item \emph{Dominance of one strategy}. If $ab\leq 0$ and at least one of the factors $a$ and $b$ is not $0$, then $x^*\notin (0,1)$, and the dynamics will lead either to $x=0$ or to $x=1$ depending on the sign of $\dot{x}$. This is what happens in prisoner's dilemmas (PD, defection dominates) and harmony games (HG, cooperation dominates).

\begin{center}
\begin{minipage}[b]{0.5\linewidth}
\begin{picture}(160,10)(0,0)
\thicklines
\put(30,2){\circle{3}}
\put(31.5,2){\line(1,0){77}}
\put(31.5,2){\vector(1,0){40}}
\put(110,2){\circle*{3}}
\end{picture}
\end{minipage}
\end{center}

\item \emph{Bistability}. If $ab>0$ and $a,b<0$ then $x^*$ is an unstable rest point and the dynamics will lead either to $x=0$ or to $x=1$ if the initial fraction of individuals is below or above $x^*$ respectively. This is what happens in stag hunt (SH) games.

\begin{center}
\begin{minipage}[b]{0.5\linewidth}
\begin{picture}(160,10)(0,0)
\thicklines
\put(30,2){\circle*{3}}
\put(31.5,2){\line(1,0){27}}
\put(57,2){\vector(-1,0){15}}
\put(60,2){\circle{3}}
\put(63,2){\vector(1,0){25}}
\put(108.5,2){\line(-1,0){47}}
\put(110,2){\circle*{3}}
\end{picture}
\end{minipage}
\end{center}

\item \emph{Coexistence}. If $ab>0$ and $a,b>0$ then $x^*$ is a stable attractor, and whenever the initial composition contains a mixture of individuals, the dynamics will lead to stable coexistence of strategies in proportions $x^*$ for A and $1-x^*$ for B. This is what happens in snowdrift (SD) games.

\begin{center}
\begin{minipage}[b]{0.5\linewidth}
\begin{picture}(160,10)(0,0)
\thicklines
\put(30,2){\circle{3}}
\put(31.5,2){\line(1,0){27}}
\put(33,2){\vector(1,0){15}}
\put(60,2){\circle*{3}}
\put(107,2){\vector(-1,0){25}}
\put(108.5,2){\line(-1,0){47}}
\put(110,2){\circle{3}}
\end{picture}
\end{minipage}
\end{center}

\end{enumerate}

Thus, in this case in which the replicator dynamics describe the evolution of the system (most dynamics behave in a similar way in the mean field limit), the PD --and the public goods game (PGG) with strong altruism as its n-players generalisation (see \cite{hauert:2003, perc:2013} for two ways to map PGG's into two-player games)-- is the most difficult case for the evolution of cooperation, as the evolutionary outcome is dominance of defection.

\section{The prisoner's dilemma: Altruism versus free-riding and beyond}
\label{sec:PD} 
The PD game has been widely used as a mathematical metaphor representing the problem of the evolution of altruism and cooperation during the last 20 years. However, its study by theoretical scientists has been surrounded by controversy since the very beginning, as some scientists claim that other games, as the snowdrift, in which coexistence is the evolutionary outcome (see section~\ref{sec:DynamicalOutcomes}), are more appropriate to represent real interactions. The controversy has not yet been solved, as difficulties arise when trying to measure payoffs in nature, which usually does not allow to assess if the payoff ordering is that of a PD or of a snowdrift \cite{turner:1999,turner:2003,hauert:2004,doebeli:2005}. I show in section~\ref{sec:onlyPDandHG} below a special feature that allows for a clarification of this problem in some contexts: If individual behaviours produce a fixed fitness variation on the actor, a fixed fitness variation on the receiver of the act, and fitness is additive, only PD structures or harmony games emerge, being cooperation the non-trivial solution only for the PD. However, I also show a limitation of the PD: the widely spread belief that any PD structure of the interactions involves altruists and non-altruists is wrong. As I show in sec.\ref{sec:defectiveAlt}, two altruists interacting together may also define a PD, in which they differ in their cooperativeness degree.

\subsection{Direct interactions and additive payoffs lead either to prisoner's dilemmas or to harmony games}
\label{sec:onlyPDandHG}
Suppose that in a habitat there are two interacting species, or in a population two different behavioural types; let's call them A and B. Suppose also that during an interaction individual A produces a fitness change $A_s$ on itself and $A_r$ on the recipient, and B produces $B_s$ and $B_r$ on itself and on the co-player respectively. The interaction matrix is
\begin{equation}
\label{eq:PDmatrix}
\begin{tabular}{l|cc}
 & A & B \\
\hline
A & $R = A_s + A_r$ & $S = A_s + B_r$ \\
B & $T = B_s + A_r$ & $P = B_s + B_r$
\end{tabular} 
\end{equation}

Let us assume in the following that individual A is the cooperative individual, and B the non-cooperative one, also called defector. Then, the first requirement in the definition of cooperative behaviours (see section~\ref{sec:reldef}) --it is better to interact with the most cooperative individual-- turns into $T,R > P,S$, and the second requirement --absolute cooperative individuals do not reduce the recipient's fitness-- into $A_r\geq 0$. Defining $\Delta_s = A_s - B_s$ (relative selfishness degree of A), and $\Delta_r = A_r - B_r$ (relative cooperativeness degree of A), the two requirements for the definition of a behaviour as cooperative reduce to 
\begin{equation}
\label{eq:Cgames}
\begin{tabular}{l}
$\Delta_r > Abs(\Delta_s)$ \\
$A_r \geq 0$.
\end{tabular}
\end{equation}

The first of equations~\eqref{eq:Cgames} implies that, for individual A to be a relative cooperator and B a relative non-cooperator (usually called defector), the relative cooperativeness degree of A must be bigger than the absolute value of its relative selfishness degree. The second equation requires the action of A on B to have neutral or positive effects in order to call A an absolute cooperator -- note that absolute refers to the chosen framework. As we will see below in section~\ref{sec:defectiveAlt}, A being an absolute cooperator does not imply that B is a non-cooperator; they may perfectly be both absolute cooperators, and still determine a PD. For that reason, in the following cooperative and defective individuals should be interpreted according to the relative scale defined by equations~\eqref{eq:Cgames}.

There are four possible games that are consistent with equations~\eqref{eq:Cgames} (see section~\ref{sec:DynamicalOutcomes}), and thus one of the interacting individuals is regarded as more cooperative than the other:
\begin{equation}
\label{eq:PD}
\begin{tabular}{llll}
Game & Payoff ordering & $\to$ & Requirement \\
Prisoner's Dilemma (PD) & $T \geq R > P \geq S$ & $\to$ & $\Delta_s \leq 0$ \\
Harmony Game (HG) & $R \geq T > S \geq P$ & $\to$ & $\Delta_s \geq 0$\\
Snowdrift (SD) & $T \geq R > S \geq P$ & $\to$ & $\Delta_s = 0$ \\
Stag Hunt (SH) & $R \geq T > P \geq S$ & $\to$ & $\Delta_s = 0$
\end{tabular}
\end{equation}

For a PD, being A the cooperative individual, the relative selfishness degree of A is negative or zero, i.e. $\Delta_s \leq 0$. This means that it is better to be a defector (relative non-cooperator) than a cooperator; as $A_s \leq B_s$, changing strategy from A (cooperate) to B (defect) in any interaction increases ones own payoff in a quantity $\Delta_s$, while getting the same payoff from the co-player. Thus, Darwinian selection of the fittest promotes defection, and as for a PD selfishness is anti-correlated with cooperativeness, evolution leads to populations of the most selfish individuals, where everyone earns a payoff $P$, even if populations of relative cooperative individuals do have a higher mean payoff, $R>P$. Note that whenever $P=0$ we face the so called tragedy of the commons \cite{hardin:1968}, i.e. the exhaustion of common goods due to selfishness.

If one imposes the payoff ordering for a harmony game, then $\Delta_s \geq 0$, i.e. $A_s \geq B_s$. In this case cooperativeness is correlated with selfishness, i.e. cooperative individuals (A) are also the more selfish, and any individual increases its self-payoff in a quantity $\Delta_s$ in any interaction by changing to cooperate instead of defecting, independently of the co-player's strategy. Cooperators are thus favoured by natural selection and their evolution does not represent a dilemma, as self-maximisation of payoffs equals mean population-payoff maximisation. This kind of situations, in which selfishness and cooperativeness are (positively) correlated, are the situations to which Adam Smith referred when he spoke about an \emph{invisible hand} at work, by which individuals increase the common good by their selfish motifs \cite{smith:1776}. This is however not true for PD's, when defection is promoted by natural selection. 

For the snowdrift (SD) and the stag hunt (SH) games, the payoff ordering imposes $A_s = B_s$, which means that, if payoffs are assumed to be additive and interactions direct, both games only exist at the boundary between PD and HG regions, and might be seen as limit cases of them. It is not possible to have a SD or SH with fixed additive fitness variations due to direct interactions; only PD and HG structures emerge in this case, i.e. all possible combinations of behaviours present in Fig.~\ref{fig:diagramanp} give rise to PD's or HG's.

\subsection{Cooperative and defective altruists in a prisoner's dilemma and the paradox of goodness}
\label{sec:defectiveAlt}
The payoff ordering in equation~\eqref{eq:PD} implies that whenever there are two altruistic behaviours interacting together, i.e. $A_s = -c_{a}$, $A_r = b_{a}$, $B_s = -c_{b}$, $B_r = b_{b}$, and both behaviours fulfil a PD, i.e. $ c_{a} \geq c_{b}$, $b_{a} \geq b_{b} + c_{a} - c_{b}$, then A is regarded as cooperator and B as defector. This has led to the false and widely spread belief that PD structures always include absolute cooperative and non-cooperative individuals. According to the definition of cooperation given in section \ref{sec:reldef} and formalised in equations~\eqref{eq:Cgames}, altruistic behaviours are always absolute cooperative behaviours when compared to the neutral individuals defining the origin of the reference system (see Fig.~\ref{fig:diagramanp}), and thus, although one of them is more cooperative and less selfish than the other, two absolute cooperative behaviours may determine a PD. 

Note that this leads to the paradox of goodness: Whenever one individual is the most cooperative and less selfish in a population, any interaction in which he is involved will be a PD, where he is regarded as cooperator irrespectively of any other strategies determining HG's or PD's. Being this so, for the most cooperative and less selfish individual, his interaction environment is always a PD in which any other individual is subject to the selfish temptation to defect when interacting with him, irrespective of being a fully altruistic population, or any other. 

The paradox of goodness represents a strong version of the paradox of cooperation. Cooperation is promoted in HG's, when the invisible hand is at work by means of the positive correlation of selfishness and cooperativeness. However, even in a population in which the invisible hand is at work, goodness (higher cooperativeness, lower selfishness) is never promoted by natural selection alone. It seems thus necessary to address the behavioural roots that lead to PD's, and to find mechanisms that promote goodness. In section~\ref{sec:TwoPDs} I concern about both: I show first that whenever an arbitrary behaviour is used as reference, two fundamentally different PD's arise, one in which cooperators increase social goods above the reference value, and another in which defectors decrease them below zero, and the reference strategy (which may still be regarded as a free-rider) represents goodness. Then, I review some of the mechanisms that promote altruism and focus on their capacity to promote goodness. 

\subsection{Two prisoner's dilemmas: Altruists, free-riders and parasites.}
\label{sec:TwoPDs}
The PD may represent two different situations, one related to social goods formed by cooperation, another to social goods created by refraining from conflict \cite{rankin:2007}. One may prove this using a free-rider (passive, neutral individual) to define the null point of the reference system (Fig.~\ref{fig:diagramanp}); then, altruists that pay a cost $c_a$ to give a benefit $b_a>c_a$ to the co-player are more cooperative and less selfish, creating social goods that increase the mean population fitness at a cost to themselves, while if compared to parasites that pay a cost $c_p$ to parasite a fitness amount $b_p$ from the co-player, free-riders are cooperative individuals (relative and absolute). In this case the social good is the non-competitive environment created by free-riders, while parasites decrease the mean population fitness. 
In short:
\begin{equation}
\label{def:strategies}
\begin{tabular}{ll}
Altruists: & Pay $c_a$, give $b_a$ to the co-player. \\
Free-riders: & Receive the action of the co-player. \\
Parasites: & Pay $c_p$, parasite $b_p$ from the co-player.
\end{tabular}
\end{equation}

The interaction matrices determined by altruists and free-riders, and by free-riders and parasites
\begin{equation}
\label{eq:PDmatrixes}
\begin{tabular}{ll}
\begin{tabular}{l|cc}
(a) & C & D \\
\hline
C (altruist) & $b_a - c_a$ & $- c_a$ \\
D (free-rider) & $b_a$ & $0$
\end{tabular} 
&
\begin{tabular}{l|cc}
(b) & C & D \\
\hline
C (free-rider) & $0$ & $-b_p$ \\
D (parasite) & $b_p - c_p$ & $-c_p$
\end{tabular}
\end{tabular}
\end{equation}
determine a PD according to equation~\eqref{eq:PD} whenever 
\begin{equation}
\label{eq:PDcondition}
b_i > c_i > 0
\end{equation}

Although in recent years some evolutionary game theorists have carried out work on the study of cooperation using generic payoffs $T,R,P,S$, most work on the evolution of cooperation has focused on the study of altruists versus free-riders. Specifically, most mechanisms found for the evolution of cooperation refer to the evolution of altruism. Thus, we may ask ourselves if this rules work in the free-rider versus parasite case.

\subsection{Mechanisms promoting cooperation in the altruist versus free-rider dilemma} 
\label{sec:mechanisms}
The study of the PD has led to the discovery of a set of mechanisms allowing for the survival and expansion of altruism \cite{nowak:2006b,lehmann:2006}, which might be classified in two groups: structural and behavioural mechanisms. These mechanisms usually apply with no (or slight) change to the public goods game (PGG) as the N-players generalization of the PD --note however than in a usual PGG cooperative individuals get self-benefits, which does not happen in the PD. 

At this point it is important to remark that all the mechanisms for the evolution of cooperation rely on the same basic property: \emph{assortment}. If the assortment between cooperative individuals is high enough, i.e. the number of cooperator-cooperator interactions is high enough as to ensure that the benefits of cooperation are higher for cooperators than for defectors, then cooperation is favoured by the natural selection process and has higher chances to take over the entire population than defection (see \cite{fletcher:2009} for a discussion of assortment in PGG's).

Structural mechanisms ensuring high enough levels of assortment rely on the existence of some factor which does not depend directly on the behaviour of the individuals, and which roots on spatio-temporal or matter-energy factors. This mechanisms include:
\begin{itemize}
\item \emph{kin selection} \cite{hamilton:1964a, hamilton:1964b, wolf:2011}, which allows for the evolution of cooperation whenever altruistic behaviours are linked to genes, as explained in the introduction.
\item \emph{network structures}, as space or interaction networks, which depending on the properties of the network may allow for the evolution of cooperation both for prisoner's dilemmas \cite{ohtsuki:2006,pacheco:2006,poncela:2008,gomez-gardenes:2007,roca:2009} and public goods games \cite{santos:2008,perc:2013}.
\item \emph{multilevel or group selection} \cite{traulsen:2006, lehmann:2006b}, which happens whenever there are groups of individuals, the individuals interact according to a prisoner's dilemma only with individuals of their group, and there is also competition and selection (birth-death process) at the group level. In some situations this allows for cooperation to thrive \cite{traulsen:2006}.
\item \emph{green-beards} \cite{dawkins:1989,smukalla:2008,sigmund:2009b,gardner:2010}, which happens when the altruistic behaviour is genetically coded and preferentially directed towards individuals carrying the altruistic trait. The green-beard effect has been found in nature \cite{gardner:2010}, where green beard--like genes code for cell adhesion \cite{queller:2003,smukalla:2008,gardner:2010}.
\item \emph{linking payoffs to ecological variables}, which has been proven to allow for the evolution of cooperation in the competition for oviposition sites in insects \cite{mesterton-gibbons:1991}, and in general situations in which a limiting resource constrains the replication and interaction capacity of the individuals \cite{requejo:2011,requejo:2012,requejo:2013,requejo:2012b}. Including mobility and variable quality of habitats has also been proven to increase assortment as to allow for the evolution of altruism \cite{pepper:2002,pepper:2007}. Finally, ecological dynamics including variable population sizes have been recently studied in a PGG \cite{hauert:2006a,wakano:2007,hauert:2008,wakano:2009,wakano:2011} showing that it allows for cooperation to thrive; however, this mechanism fails for the PD \cite{hauert:2006a}.
\end{itemize}
In addition to the previous mechanisms, it is worth to analyse the case of weak altruism. In the PGG, as the benefits created by cooperative individuals are redistributed within the entire interaction group, cooperative individuals get a self-benefit for their own cooperative action. This makes cooperative individuals perceive always an extra cooperator (itself) in their environment. In the case in which the self-benefit of cooperation is enough as to compensate the cost paid, the structure of random interactions allows for the evolution of altruism in well-mixed populations. However, as individuals are not really experiencing any net cost --although they still increase the other players' fitness more than theirs-- this case is usually called weak altruism (See \cite{wilson:1990, fletcher:2009} for nice reviews). Furthermore, this case cannot be mapped into an altruist vs. free-rider PD whenever interaction groups consist of two individuals, as in the PD altruists experience a net cost (see section 2 in \cite{hauert:2003} for a nice explanation), being thus the case of weak altruism less restrictive than the PD, which corresponds to strong altruism (note that weak altruism corresponds to selfish cooperation, as shown in Fig.\ref{fig:diagramanp}.

Behavioural mechanisms imply the addition of new behavioural types, which may require the use of higher cognitive abilities, as memory or recognition capacity. Behavioural mechanisms found to promote altruism in a well-mixed prisoner's dilemma, where every individual interacts with any other, include
\begin{itemize}
\item \emph{reciprocity \cite{trivers:1971,axelrod:1981,nowak:1998,nowak:2006a,sigmund:2010} --direct \cite{axelrod:1981}, indirect\cite{nowak:1998} and generalized \cite{yamagishi:2000,pfeiffer:2005}--}, where individuals choose to cooperate or not according to some previous information about previous interactions, as for instance if the co-player cooperated with you previously (direct reciprocity \cite{axelrod:1981,nowak:2006a,sigmund:2010}), if he cooperated previously with others (indirect reciprocity \cite{nowak:1998,sigmund:2010}), or if there were previous cooperative interactions in the group \cite{yamagishi:2000,pfeiffer:2005}. The most famous example of a behaviour that allows for the evolution of altruism through direct reciprocity is Tit For Tat, a strategy which cooperates the first time it interacts with someone, and then it simply imitates the behaviour displayed by the co-player in their last interaction together. This very simple behaviour turned out to be the surprising winner of a series of computer tournaments \cite{axelrod:1981}, showing that one time step memory is enough to promote cooperation whenever the game is an iterated prisoner's dilemma, i.e. played repeatedly between the same two players for many rounds.
\item \emph{punishment and reward} \cite{henrich:2006,sigmund:2001,brandt:2003,fehr:2002,fehr:2002b, hauert:2007,boyd:2010,fehr:2003b,szolnoki:2010,szolnoki:2011,szolnoki:2012}, when altruistic individuals may choose to pay some extra cost in order to punish (impose a cost on) a free-rider partner with whom they interacted, or to reward other altruistic individuals. Although the relative importance of both mechanisms is debated, it is generally accepted that they are able to increase cooperation levels.
\item \emph{similarity donation}, which happens whenever individuals choose to cooperate if they find a certain level of similarity between them and their interaction partners \cite{riolo:2001,axelrod:2004,jansen:2006,
traulsen:2003,traulsen:2007,antal:2009}. This mechanism is similar to a non-genetically determined green beard, explained above. If there is a strong enough positive correlation between altruistic behaviours and phenotype (observable characteristic) altruism is favoured by the evolutionary process and will increase.
\end{itemize}  
In addition to reward and punishment, behavioural mechanisms promoting cooperation in the public goods game (PGG) include
\begin{itemize}
\item \emph{loner strategies} \cite{hauert:2002a,hauert:2002b,hauert:2005}, which do not play the public goods game and get some benefit on their own. The inclusion of loners in addition to cooperators and defectors allows for neutrally stable cycles in the absence of mutations, and for stable coexistence with cooperative and defective individuals whenever mutations between strategies happen.
\item \emph{destructive strategies} \cite{arenas:2011,requejo:2012c}, termed Jokers, i.e. individuals who do not play the game and damage any public goods game participant. This behaviour allows for robust evolutionary cycles of cooperation, defection and destruction, even in the presence of mutations.
\end{itemize}

All previous mechanisms enable the evolution of altruism, either promoting the invasion of the entire population by altruistic individuals, or allowing for their survival in a coexistence state with free-riders. 

In the biological literature mechanisms are often classified in a different way, taking into account the fact that individuals possess the ability to manage many of the previous mechanisms inherently, by their own inherited capacities and structural form. In this way, the basic mechanisms are usually listed as kin discrimination (\cite{hamilton:1964a,hamilton:1964b}), population viscosity (also called limited dispersal, \cite{hamilton:1964a,hamilton:1964b}), enforcing mechanisms (including the conditional behaviours above) and group augmentation (\cite{kokko:2001}). This classification is in my opinion complementary to the one given above. While the first tries to disentangle the effect of each particual feature that we find in nature, the latter classification already puts together the pieces which are found in animal species, as for limited dispersal, which accounts for spatial structure and mobility of individuals. For nice reviews from the biological perspective read \cite{lehmann:2006, west:2007, west:2010}.

\subsection{Mechanisms promoting cooperation in the parasite versus free-rider dilemma}

The mechanisms of kin selection, multilevel selection, network structures and reciprocity require for the evolution of altruism in populations of altruists and free-riders \cite{nowak:2006b} that
\begin{equation}
\label{eq:RuleAlt}
q > c_a/b_a
\end{equation}
where $q$ is a constant related to the mechanism. For direct reciprocity $q$ is the probability of playing a next round with the same player, for indirect reciprocity it is the probability of knowing the reputation of the other individual, and for kin selection $q=r$ is the genetic relatedness. 

Whenever equation~\eqref{eq:RuleAlt} is fulfilled cooperation is --at least-- evolutionary stable\cite{nowak:2006b, taylor:2007}, i.e. altruists resist invasion attempts by free-riders, but, do the mechanisms for the evolution of altruism work for the evolution of non-aggressive societies, i.e. the free-rider versus parasite case? In order to answer to this question note that, if the evolutionary dynamics of two situations are identical, the dynamical result must be the same. In appendix~\ref{sec:DynamicalEquality} the conditions for such dynamical equality are derived for the case in which the evolutionary dynamics depends either on the difference between individual payoffs, or between individual and mean population payoffs, which applied to the altruist versus free-rider and free-rider versus parasite cases (equations~\eqref{eq:PDmatrixes}) result in 
\begin{equation}
\label{eq:constraint1}
b_a = b_p
\end{equation}
\begin{equation}
\label{eq:constraint2}
k = \Delta b_a = c_p.
\end{equation}
where $\Delta b_a = b_a - c_a$.

Using the previous equations, we may rewrite equation~\eqref{eq:RuleAlt} in terms of the parameters referring to the parasite versus free-rider case. Then, the mechanisms for the evolution of non-aggressive free-riding societies in which individuals have the temptation to parasite other individuals require
\begin{equation}
\label{eq:rule}
q > 1 - c_p/b_p.
\end{equation}
Thus, cooperation is enhanced whenever the costs for defecting increase or its benefits decrease. 

Mechanisms for the evolution of cooperation based on structural properties --group selection, kin selection, network reciprocity-- work for the free-rider--parasite dilemma without modification. However, some of the mechanisms for the promotion of altruism require conditional behaviours, i.e. altruists may decide not to pay the cost in relation to some previous information about the co-player, and thus yield no benefit to the free-riders. For reciprocity mechanisms to work in the ``parasite dilemma'', when the active individual carrying out the action, and thus paying the cost, is the parasite, free-riders would have to act as a parasite when interacting with such individuals. This would increase the assortment between free-riders \cite{price:1970, hamilton:1970, fletcher:2009}, giving them the opportunity to enjoy a non-competitive environment, and reject parasites in it. It seems thus plausible that punishment directed towards parasites, which might be seen as reciprocating a parasitic act, evolved in nature parallel to emerging conflict-avoiding animal groups or societies. Indeed, it has been observed in nature that punishment of thieves (parasites) happens much more frequent than punishment directed towards lazy but non-aggressive individuals \cite{jensen:2010}.

Note also that equation~\eqref{eq:constraint2} requires $k$ to equal the synergistic benefit $\Delta b_s$ produced by the altruistic action, and the cost of the selfish act. This excludes the possibility of having the same situation in the altruist and parasite cases if $k=0$, as according to equation~\eqref{eq:PDcondition} $c_p>0$. Thus, even if the dynamics are identical, they are fundamentally different, one representing social goods formed by restraining from conflict, i.e. not decreasing other individuals fitness (free-riders versus parasite), and the second representing goods formed by cooperation (altruist versus free-riders), which increase the mean population fitness.

Every time altruists interact, they produce a benefit which is bigger than the cost paid, i.e. create a positive synergistic benefit. However, two parasites interacting together create a net cost, i.e. a negative synergy. In order to have the same evolutionary dynamics in both cases, the negative synergistic effect has to be compensated with a higher baseline fitness (equation~\eqref{eq:constraint2}). This suggests that, although in a PD cooperators always do worst than defectors, populations of altruists (active cooperators), might have some advantage over parasites (active defectors) due to the fact that the first create positive synergistic effects while parasites create a negative synergy. 

As I show in the next section, in well-mixed populations of altruists, free-riders, and parasites, the combination between the first two allows for their survival in higher levels than predicted by the mutation terms, providing a first step towards the emergence of cooperation.

\subsection{Generalised prisoner's dilemma: Altruist--free-riding assortment to survive parasitism}
\label{sec:coexistence}



The altruist and parasite strategies are defined in reference to a passive or neutral one, called free-rider. The generalised PD matrix including interactions between the three strategies is in this case
\begin{equation}
\label{eq:PDmatrixGen}
\begin{tabular}{l|ccc}
 & Altruist & Neutral & Parasite \\
\hline
Altruist & $b_a - c_a$ & $-c_a$ & $-b_p -c_a$ \\
Neutral & $b_a$ & $0$ & $-b_p$ \\
Parasite & $b_a +b_p -c_p$ & $b_p-c_p$ & $-c_p$
\end{tabular} 
\end{equation}
Note that any two strategies in this matrix determine a PD if $b_i>c_i>0$.

In any mixed population of two strategies determining a PD, defectors perform better than cooperators, i.e. have higher fitness, and thus the final population will consist only of defective individuals. Let us see what happens when the three strategies are mixed in the population. For this purpose I will assume that the replicator dynamics hold, i.e. the variation of $x_i$, the fraction of $i$ individuals in the population, follows the equation (let's recall it from Eq.~\eqref{eq:replicator0}),
\begin{equation}
\label{eq:replicator}
\frac{dx_i}{dt}=\dot{x_i}=x_i(f_i-\overline{f})
\end{equation}
where $f_i$ is the payoff of strategy $i$ and $\overline{f}$ is the mean population payoff.

The payoffs for altruists, free-riders (neutral individuals) and parasites in a well-mixed population with proportions of individuals $x_i$, $i = a, n$ or $p$, are
\begin{equation}
\begin{tabular}{l}
$f_a=1-s+s [x_a b_a-x_p b_p-c_a]$ \\
$f_n=1-s+s [x_a b_a-x_p b_p]$ \\
$f_p=1-s+s [x_a (b_a+b_p)+x_n b_p-c_p]$.
\end{tabular}
\end{equation}
where the selection strength $s$ accounts for the relative effect of the interactions on the fitness of individuals (for $s=0$ the game is irrelevant; for $s=1$ the game determines the entire fitness \cite{wild:2007}). 

\begin{figure}
\begin{center}
\includegraphics[width=85mm]{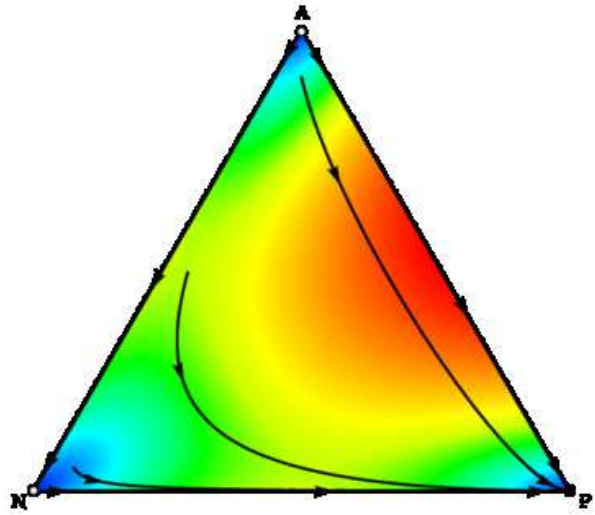}
\caption{Replicator dynamics in well-mixed populations (mean field limit) of altruists (A), free-riders (N) and parasites (P) without mutations. The simplex represents the fractions of individuals in the population; it shows the $x_a + x_n + x_p = 1$ plane in the three dimensional space determined by the fractions of strategies: The corners represent homogeneous populations of the corresponding behaviour, while any other point represents mixed states. Colours correspond to different speeds $dx_i/dt$: red for the fastest, blue for the slowest. The evolutionary dynamics leads to homogeneous populations of parasites, the dominant strategy, where the fitness of any individual is $-c_p<0$. This represents the worst possible outcome, as homogeneous populations of free-riders possess null fitness, and populations of altruists have a positive fitness equal to $b_a-c_a>0$ (note that these quantities might be seen as variations of a baseline fitness). Parameters: Altruist and parasite benefits $b_a = b_p = 2$ and costs $c_a = c_p = 1$; selection strength $s=1$. Images obtained using a modified version of the Dynamo Package \cite{dynamopackage:2011}. \label{fig:simplexnomut}}
\end{center}
\end{figure}

As $x_a+x_n+x_p = 1$, we can describe the system dynamics with two replicator equations. We will use those for altruists and parasites. The replicator equations are
\begin{equation}
\label{eq:repeqs}
\begin{tabular}{l}
$\dot{x_a} = s x_a [(b_a x_a-c_p) x_p-(1-x_a) c_a]$,\\
$\dot{x_p} = s x_p [x_a (c_a+b_a (x_p-1))+c_p (x_p-1)-b_p (x_a+x_p-1)]$,
\end{tabular}
\end{equation}
and the rest points $(x_a, x_p)$ for such dynamics are (0,0), (0,1), (1,0), corresponding to homogeneous populations of one of the strategies. However, as it can be observed in Fig.~\ref{fig:simplexnomut}, the dynamics leads to populations of parasites, the most selfish and less cooperative strategy, whenever the initial conditions include individuals of all types. Note that the selection strength only introduces a time scale change to the previous equations. However, for the replicator dynamics with mutations studied below, such selection strength is mandatory in order to ensure positive payoffs.

\begin{figure*}
\begin{center}$
\begin{tabular}{lll}
(a) & (b) & (c) \\
\includegraphics[width=50mm]{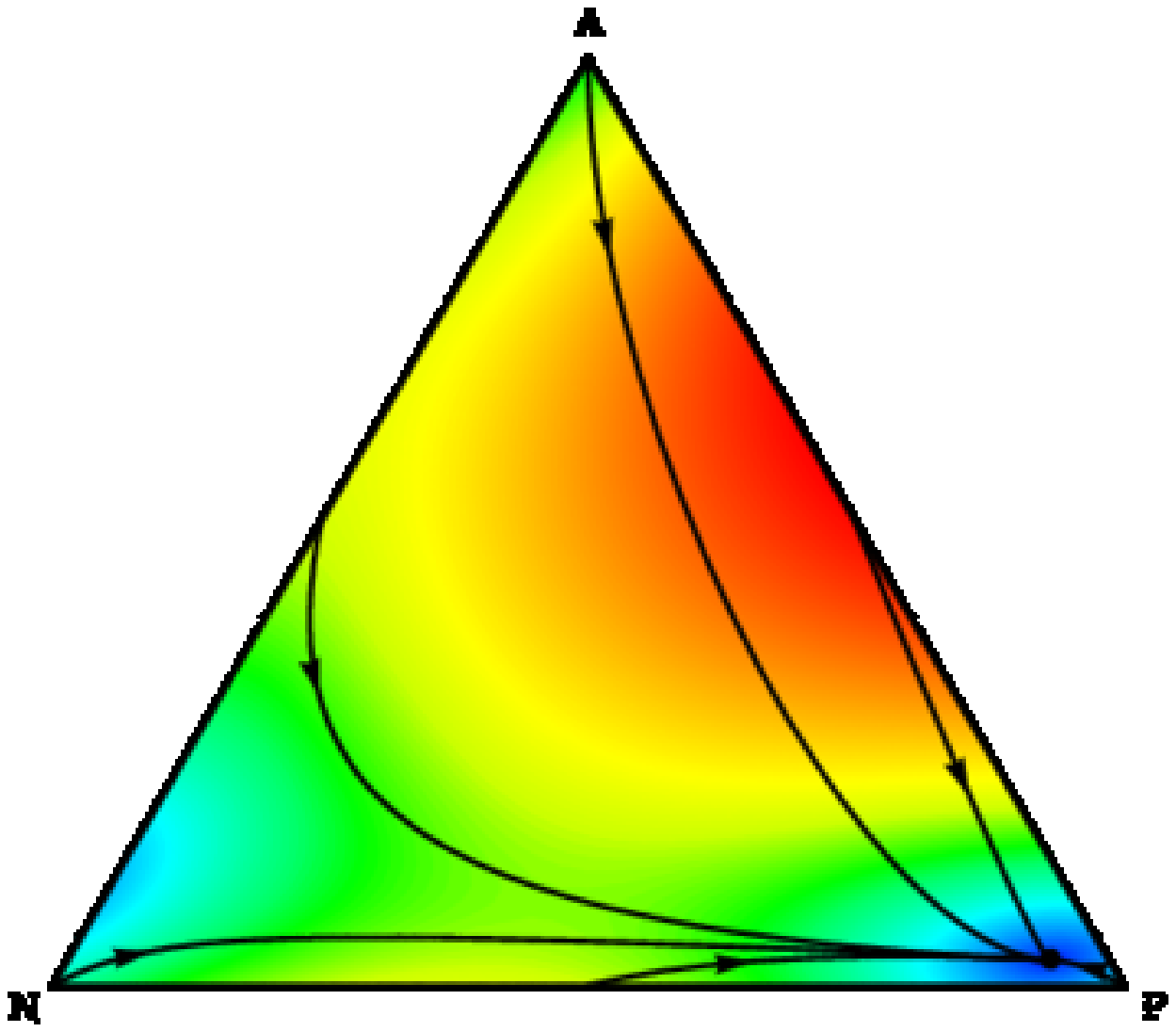} &
\includegraphics[width=50mm]{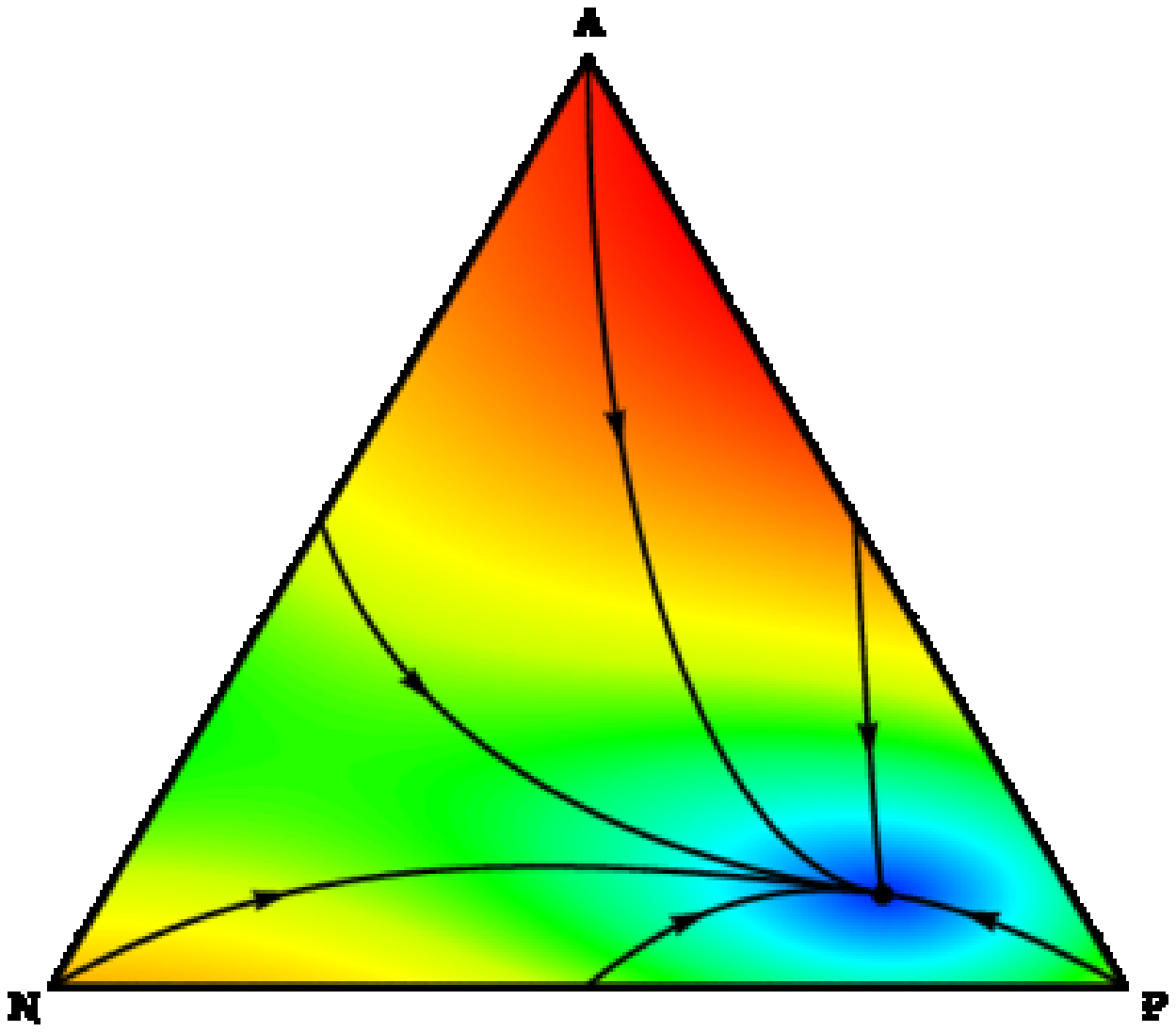} &
\includegraphics[width=50mm]{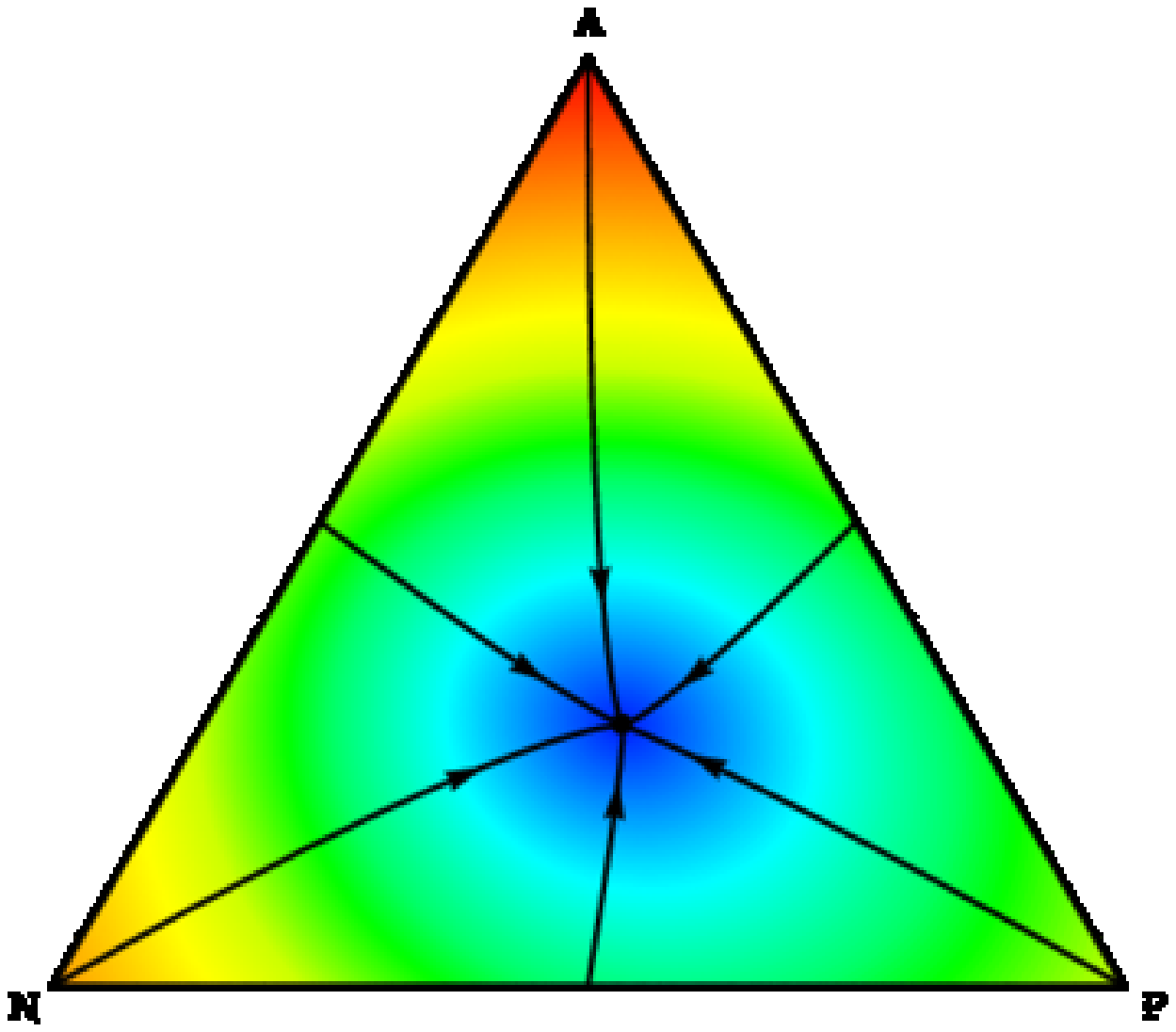} \\
(d) & (e) & (f) \\
\includegraphics[width=50mm]{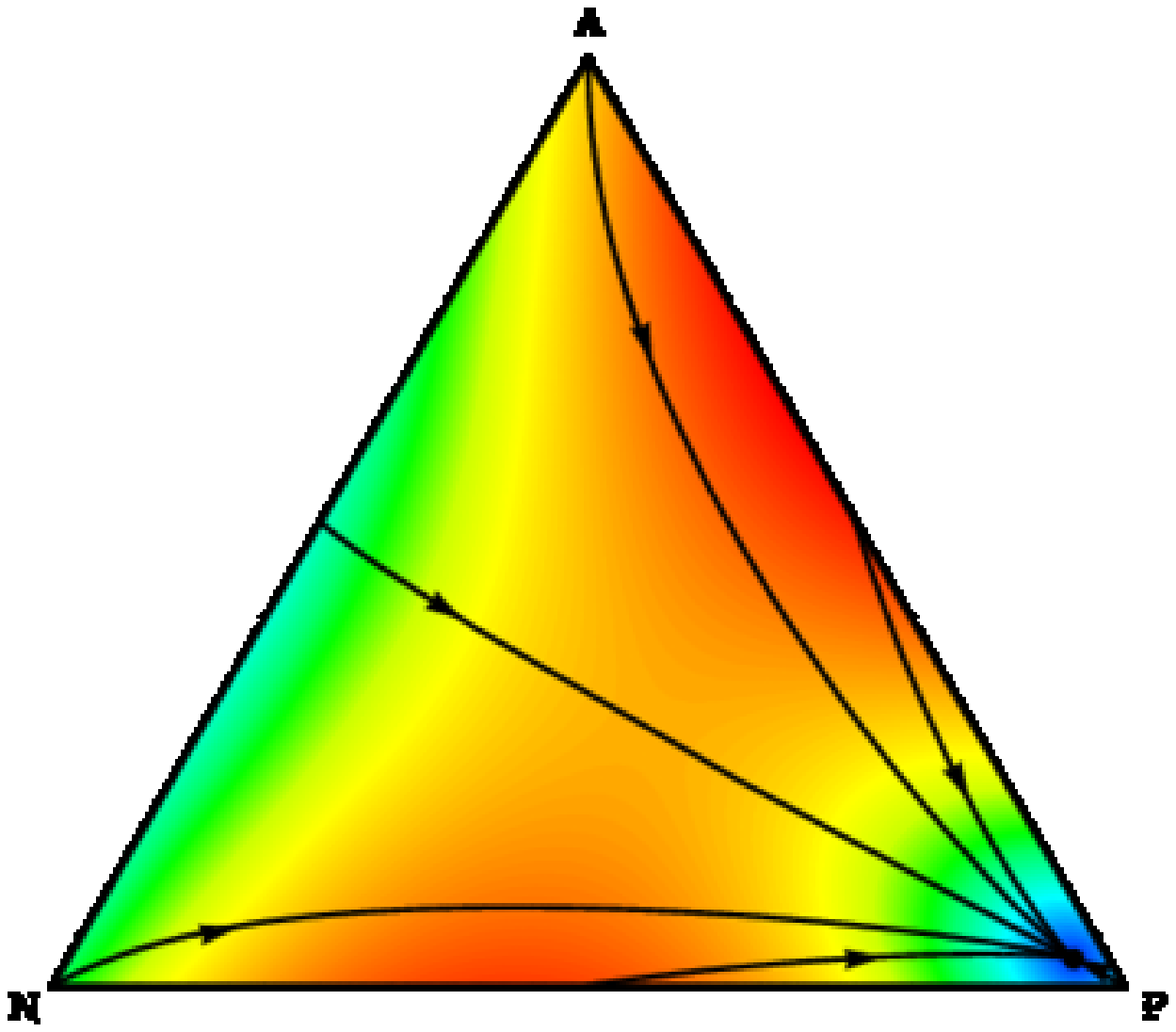} &
\includegraphics[width=50mm]{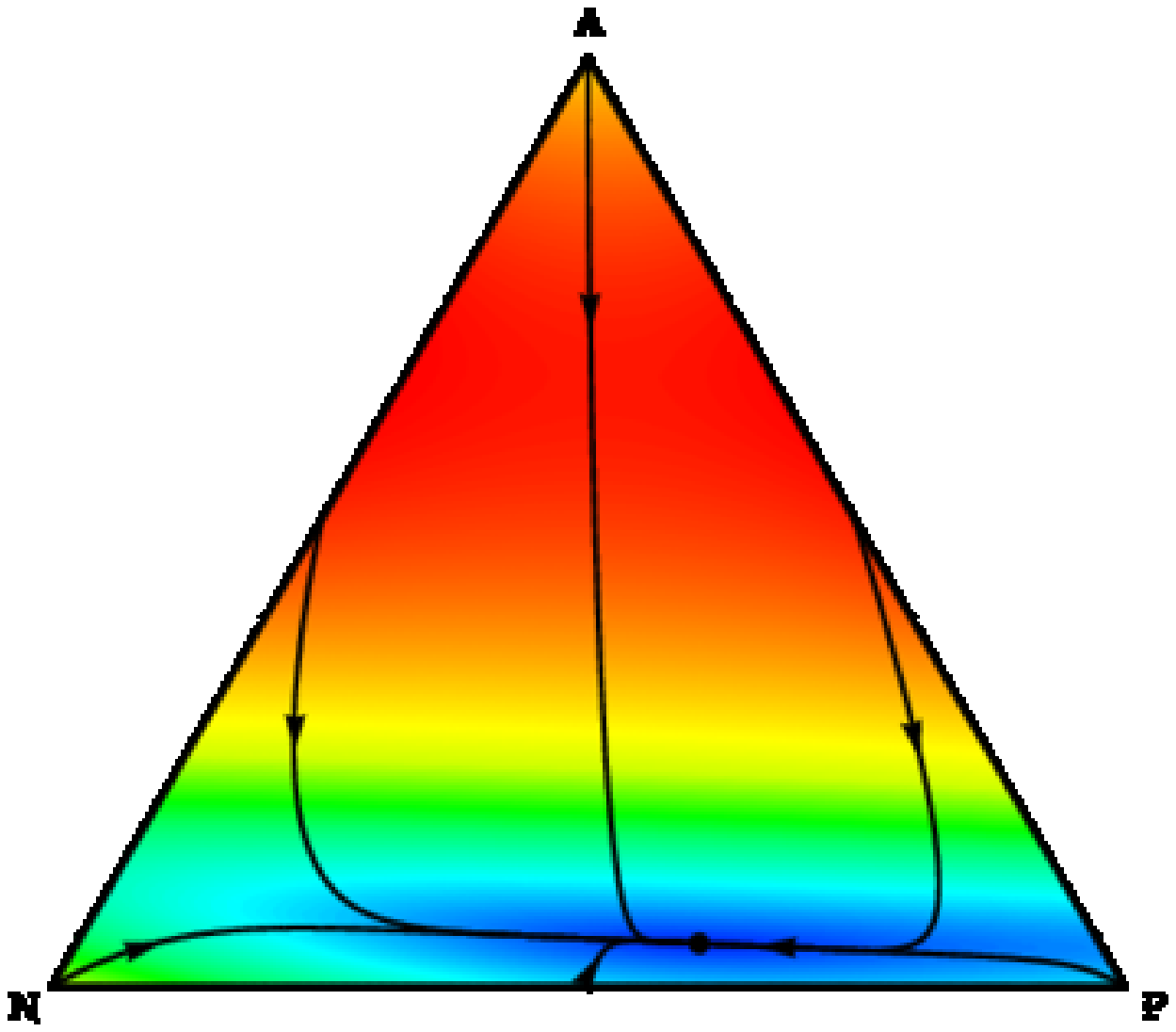} &
\includegraphics[width=50mm]{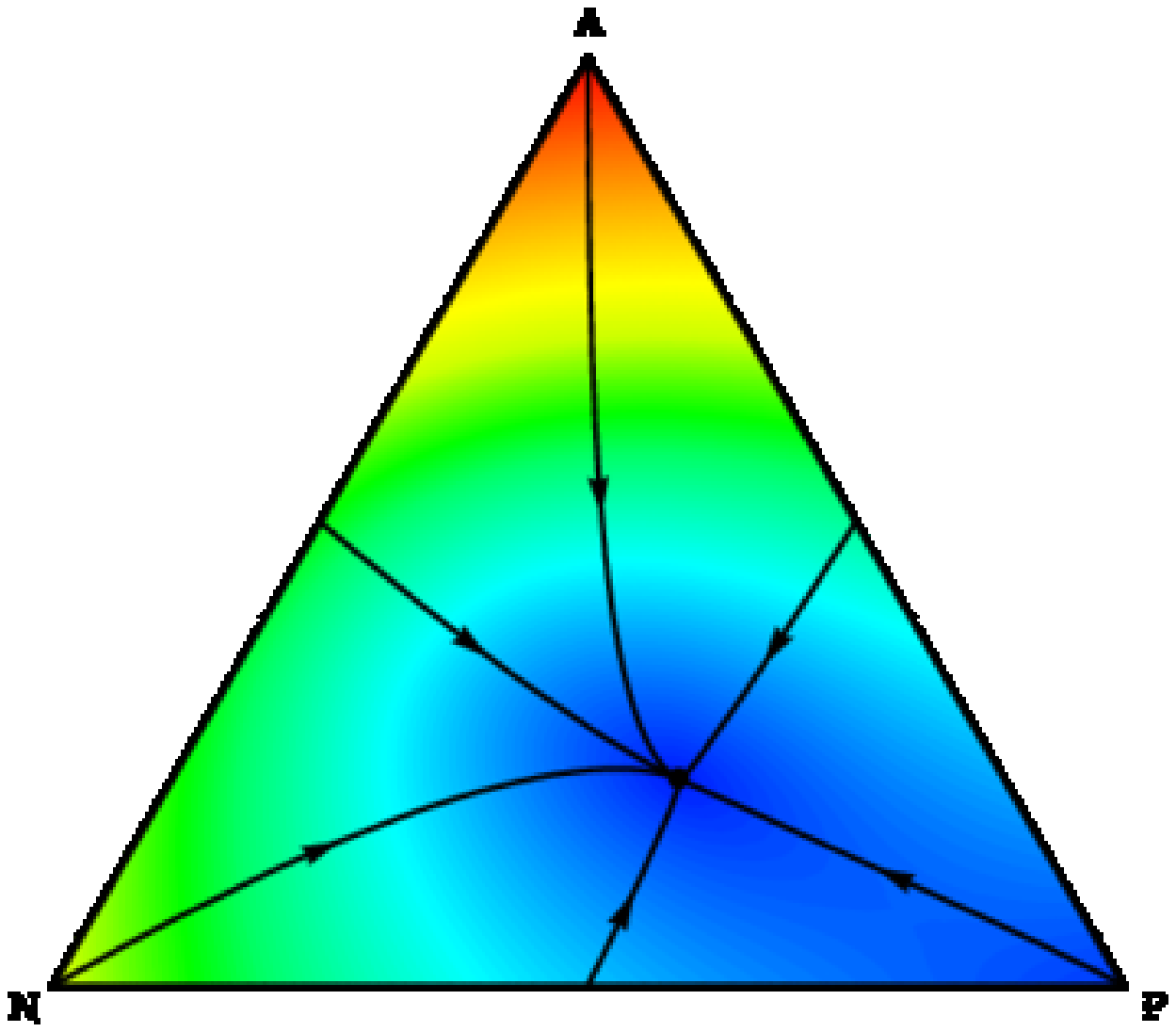}
\end{tabular}$
\caption{Replicator-mutator \ref{eq:repmut} dynamics in well-mixed populations of altruists (A), free-riders (N) and parasites (P). Unexpectedly, altruist and free-rider individuals survive in higher proportions than expected by the mutation rate. Increasing the mutation rate increases the fractions of altruists and free-riders in the fixed point, as shown in (a)-(c). Decreasing altruists costs and increasing parasites costs favours altruism, as shown in (d)-(f), which have all other parameters equal to (a) to allow for a comparison. Parameters: $b_a = b_p = 2$, $s=1/(1+c_a+b_p)$, maximum selection strength which ensures non-negative payoffs, (a)-(c) $c_a = c_p=1$, (a) $\mu=0.05$, (b) $\mu=0.15$, (c) $\mu=0.45$; (d)-(f) $\mu=0.05$, (d) $c_a=0.05$, $c_p=1$, (e) $c_a=1$, $c_p=1.95$, (f) $c_a=0.05$, $c_p=1.95$.\label{fig:simplexmut}}
\end{center}
\end{figure*}

The situation in which cooperation extinguishes changes drastically if mutations appear. In order to introduce mutations into the system the replicator-mutator equation will be used \cite{page:2002}. Note however that introducing a mutation term of the form $\mu (1-3x_i)$ in Eqs.\ref{eq:repeqs} returns similar results --this is the usual choice in cultural reproduction, while the replicator-mutator equation usually refers to genetic reproduction. The replicator mutator equation is
\begin{equation}
\label{eq:repmut}
\dot{x_i} = \sum_{j=1}{n} x_j f_j({\bf x}) q_{ij} - x_i \bar{f}
\end{equation}
where $x_i$ is the fraction of $i$ individuals in the population, $q_{ij}$ is the transitions (mutation) matrix, containing the probabilities that replication of state $i$ gives rise to $j$, $f_j({\bf x})$ is the fitness of strategy $j$, which depends on the population composition, and $\bar{f}$ is the mean population fitness. In this case payoffs are necessarily positive --which can be tuned via the selection strength-- as they relate directly with the number of offspring produced instead of with frequency changes (as in cultural reproduction), which cannot be negative.

It has already been reported that, for public goods games, high mutation rates allowing for an exploration of available strategies may promote cooperation \cite{traulsen:2009}, specially in the presence of punishing and loner strategies. However, for low mutation rates and two strategies, cooperators and defectors, the promotion of cooperation is only due to mutations and in a fraction $\mu/2$ (in general, the promotion due to mutations is $\mu/n$, where $n$ is the number of strategies). In the case of a generalised prisoner's dilemma with three strategies, altruists, free-riders and parasites, if the cost associated with altruism is not too big, and the one associated with parasitism is big, the dynamics leads to coexistence of the three strategies, with fractions of altruists, free-riders and parasites approaching $1/3$ for high mutation rates (Fig.~\ref{fig:simplexmut}(c)) or extreme costs ($b_a\gg c_a, b_p \approx c_p$, Fig.~\ref{fig:simplexmut}(f)). Hence, the combination of free-riders and altruists allows for an increase of cooperative levels in the presence of parasites.


As argued before, the coexistence found cannot be explained by the mutation term alone; the presence of altruist and free-rider individuals allows for both of them to survive to the invasion of parasites, which does not happen whenever any of such strategies is mixed only with parasites. This result resembles the speciation transitions found in other models (not PD's), where evolution leads to coexistence of individuals differing in their cooperativeness levels \cite{doebeli:2004, killingback:2010}, and the effect of damaging behaviours in the promotion of cooperation \cite{arenas:2011,requejo:2012c}. Thus, although free-riders lead altruism to extinction, and parasites lead free-riders or altruists to extinction, the presence of both free-riders and altruists together provides an escape from the worst of the tragedies of the commons, i.e. populations where all individuals are parasites decreasing the fitness of any other individual, and where the mean population fitness decreases to its minimum value below zero.

\section{Competition in a world with limited resources}
We live in a world with finite resources. Such limitation, as already pointed out by Darwin and Malthus, triggers competition between individuals. Most evolutionary game theoretical models do not take this limitation of resources into account further away than assuming a constant population size. However, such limitation or resources is responsible for the tension between metabolic pathways, as well as may have played an active role in the evolutionary transitions and formation of higher selective units.

In this section it is first analysed the kind of structure determined by trade-offs between rate and yield in resources use, which have been suggested to be responsible for the transition to multicellularity. Then, the competitive exclusion principle is explained, and all previous results presented in this review are analysed by means of such principle. To finish, it is shown that a self-organizing process between resources use and population composition allows for coexistence of cooperation and defection in the first example (to my knowledge) of an exception to the competitive exclusion principle.

\subsection{Rate versus yield trade-off: Prisoner's dilemmas, harmony games and stag hunt games.}
\label{sec:RvsY}
Trade-offs between rate and yield in resources use happen in bacterial and yeast species in which different metabolic pathways allow them to process glucose by using fermentation and/or respiration. These trade-offs have been proposed to have triggered multicellularity by cell adhesion of cooperative traits \cite{pfeiffer:2001,pfeiffer:2003,koschwanez:2011}. Let us now analyse which kind of interactions structure appears whenever the two different strategies using resources, rate maximizers and yield maximizers, interact.

Rate maximizer strategies maximize the production of new individuals per unit time, e.g. the production of biomass used for reproduction per unit time. 
Yield maximizers are individuals which maximize the production of new individuals per unit resource consumed. 
Let us suppose that there is an influx of resources $r_0$ into the system which is completely depleted by one of the strategies in order to reproduce. 
A rate maximizer is able to effectively use an amount $r_r$ of such resources to reproduce, resulting in $n_r$ new individuals in a time $t_r$;
a yield maximizer uses an amount $r_y$ of resources in a time $t_y$ in order to produce $n_y$ new individuals. Let us finally assume that the cost of producing a new individual is 
\begin{equation}
\label{eq:equalcost}
k=\frac{r_x}{n_x}
\end{equation}
where $k$ is a proportionality constant between used resources and number of individuals produced and $x$ corresponds to either yield $y$ or rate $r$ maximizer strategies. Thus, the amount of wasted resources, which will be assumed not viable to be used for further reproduction, is 
\begin{equation}
\label{eq:waste}
w_x = r_0 - r_x
\end{equation}
being $w_r>w_y$, as rate maximizers use resources faster at a higher waste production rate. 

The trade-off between rate and yield maximizers can thus be captured by the following inequalities
\begin{equation}
\label{eq:tradeoff}
\begin{tabular}{l}
$n_y > n_r$ \\
$\frac{n_y}{t_y} < \frac{n_r}{t_r}$
\end{tabular}
\end{equation}
representing the first one the higher efficiency of yield maximizers, the second the faster reproduction of rate maximizers when using $r_0$ resources (this is easily proved using Eqs.~\ref{eq:equalcost} and \ref{eq:waste}). Note that this also implies not only that the time that it takes for yield maximizers to deplete the amount or resources is bigger than that of rate maximizers, i.e. $t_y>t_r$, but that it has to be at least bigger in a factor proportional to the ratio of the numbers of new yield maximizers to new rate maximizers, i.e. $t_y>(n_y/n_r) t_r$.

Now, suppose that there are two individuals for which there are $2r_0$ resources available. If those two individuals are both rate or yield maximizers, the number of new individuals produced will be the double of those produced by just one individual, i.e. $2n_x$, in a time $t_x$, producing also a double amount of wasted resources; the per-capita numbers remain unchanged. What happens if the two individuals follow different strategies? 

The total resources consumption speed is $v_x = r_0/t_x$ with $x=y,r$ for rate and yield maximizers respectively. Whenever there is an amount of resources $2r_0$ to be shared by a rate and a yield maximizer, the total consumption speed will be $v_T = v_y + v_r$, and the consumption time can be obtained $t_c = 2r_0/v_T$, which after a few calculations results in
\begin{equation}
t_c = \frac{2t_y t_r}{t_y + t_r}
\end{equation}
The amount of resource used by each individual will be $r'_x = v_x t_c$, which may be rewritten as
\begin{equation}
r'_x =2 r_0 f_x 
\end{equation}
where the fractions of resources used by each individual are
\begin{equation}
\label{eq:rvsyfractions}
\begin{tabular}{l}
$f_y = \frac{t_r}{t_y + t_r}$\\
$f_r = \frac{t_y}{t_y + t_r}$
\end{tabular}
\end{equation}
both fractions adding up to unity, $f_y + f_r = 1$, as expected. The number of new individuals produced by each strategy will be $n'_x = 2 f_x n_x$ and $n'_y = 2 f_y n_y$, and if we assume that the payoff (as a measure of fitness) of a strategy is proportional to the number of new individuals, we can write the following payoff matrix:
\begin{equation}
\label{eq:rvsymatrix}
\begin{tabular}{l|cc}
 & Y & R \\
\hline
Y & $R = n_y$ & $S = 2 f_y n_y$ \\
R & $T = 2 f_r n_r$ & $P = n_r$
\end{tabular} 
\end{equation}

The interaction matrix given by Eq.~\ref{eq:rvsymatrix} contains the information about the system, and its analysis will tell us which kind of interactions happen in the rate versus yield maximization problem. First, following Eq.~\ref{eq:tradeoff} it is straightforward that $R>P$. Furthermore, as $f_r>1/2>f_y$, it follows that $T>P$ and $R>S$. We are just left proving that $T>S$ in order to prove that the rate versus yield interactions always determine non-cooperative games, as defined in section~\ref{sec:nonCgames}. From the definition of both terms in Eq.~\ref{eq:rvsymatrix} it follows that $T>S$ is equivalent to $f_r/f_y > n_y/n_r$. Using the definition of $f_x$ given in Eq.~\ref{eq:rvsyfractions} we find that this happens whenever $t_y/t_r > n_y/n_r$, which is always true, as it is one of the conditions for the trade-off between rate and yield maximization expressed in Eq.~\ref{eq:tradeoff}. Hence, $T,R>P,S$, and the rate versus yield trade-off determines always a non-cooperative game.

We may now ask ourselves which region of the parameters space corresponds to each game, i.e. to the prisoner's dilemma, snowdrift, stag hunt and harmony games, as defined in section~\ref{sec:nonCgames}. The boundary between regions is determined by the lines resulting from $T=R$ and $P=S$. For the first boundary, $T=R$, from the definitions of $T$ and $R$ we get $2 f_r n_r = n_y$. If we define
\begin{equation}
\label{eq:taunudef}
\begin{tabular}{l}
$\tau = t_y/t_r$ \\
$\nu = n_y/n_r$
\end{tabular}
\end{equation}
where $\tau>\nu>1$ (see Eqs.~\ref{eq:tradeoff}), we obtain the boundary condition:
\begin{equation}
\label{eq:boundaryTR}
T=R \to \nu(\tau+1) - 2\tau = 0
\end{equation}
\begin{figure}
\begin{center}
\includegraphics[width=180mm]{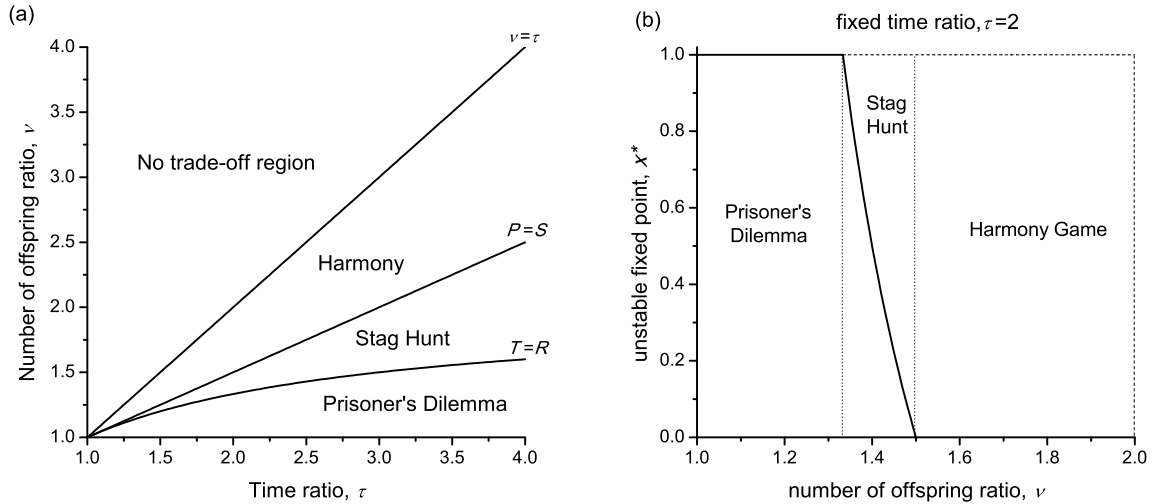}
\caption{Non-cooperative games determined by the rate versus yield trade-off. (a) The three kinds of games determined by rate versus yield trade-offs are shown as a function of the time ratio $\tau$ and number of offspring ratio $\nu$, where the trade-off condition imposes $\tau>\nu>1$. (b) The unstable fixed point $x^*$ corresponding to the replicator dynamics for a fixed $\tau=2$ is shown. For small $\nu$ we have a prisoner's dilemma and cooperators (yield maximizers) extinguish irrespective of their initial frequency; for medium $\nu$ values the interactions follow a Stag Hunt structure and cooperation extinguishes or takes over the entire population if their initial frequency is below or above $x^*$ respectively; for big $\nu$ values the interactions determine a Harmony game and defection (rate maximization) extinguishes. \label{fig:RvsY}}
\end{center}
\end{figure}
The boundary $P=S$ is equivalent to $2 f_y n_y = n_r$, which is the same as the one in the paragraph before, but with the substitution $r \leftrightarrow y$, and thus, according to Eq.~\ref{eq:taunudef}, $\tau \to \tau^{-1}$ and $\nu \to \nu^{-1}$ in Eq.~\ref{eq:boundaryTR}, resulting in:
\begin{equation}
\label{eq:boundaryPS}
P=S \to \tau - 2\nu + 1 = 0
\end{equation}
again with $\tau>\nu>1$. As shown in Fig.~\ref{fig:RvsY}, three different regions appear for the rate versus yield trade-off: a prisoner's dilemma, a stag hunt game and a harmony game. For a fixed time ratio $\tau$, the PD, SH and HG regions are found in order with increasing the number of offspring ratio $\nu$. This means that there is a region (PD) where yield maximizers (cooperation) extinguish if $\nu$ is small enough, a region of bi-stability where either yield maximizers or rate maximizers extinguish depending on their initial frequencies for medium $\nu$ values, and a third region where yield maximizers extinguish for big $\nu$. This may be illustrated assuming that the replicator dynamics drive the evolution of the system. For the replicator dynamics the unstable fixed point in the stag hunt region takes the form 
\begin{equation}
x^* = \frac{2\nu-\tau-1}{\nu+\tau-\nu \tau-1}
\end{equation}
which allows for the plot shown in Fig.~\ref{fig:RvsY}(b), in which it can be observed the effect of increasing $\nu$ for fixed $\tau$, and the appearance of the three different regions.

Note that the situation just described, and whose essence is captured by the payoff matrix given in Eq.~\ref{eq:rvsymatrix}, refers to interactions which are mediated by the environment, and not to direct interactions, as individuals are not exchanging resources, but taking them from the environment in a way which affects the co-player. Indeed, if one tries to impose the equal gains from switching condition, i.e. $T-R=P-S$, which happens for direct interactions and payoff additivity, as assumed in the section~\ref{sec:onlyPDandHG}, one obtains the condition $\nu=1$, which is opposed to the condition $\nu>1$ for an environmentally mediated rate versus yield trade-off. Indeed, the intersection point of Eqs.~\ref{eq:boundaryTR} and \ref{eq:boundaryPS}, the lines separating the three regions, happens for $\nu=1$, where the SH region disappears and only HG's and PD's exist, as for additive payoffs and direct interactions.

\subsubsection{Weak rate versus yield trade-off}

It is interesting to take into account what happens whenever the competition between rate and yield maximizers is weak, i.e. when the two strategies diverge, but not too much \cite{wild:2007}. It has been argued previously that mutations do not usually lead to big evolutionary changes all at once, but that the evolutionary process is actually a process in which the accumulation of many small mutations drives the subsequent observable effects. Thus, we may take into consideration what happens whenever there is a stable population and a random mutation drives the appearance of a new behaviour which determines a rate versus yield trade-off, but which differs only slightly from the main behaviour in the population, i.e. whose $\nu$ and $\tau$ values approach one.

\begin{figure}
\begin{center}
\includegraphics[width=180mm]{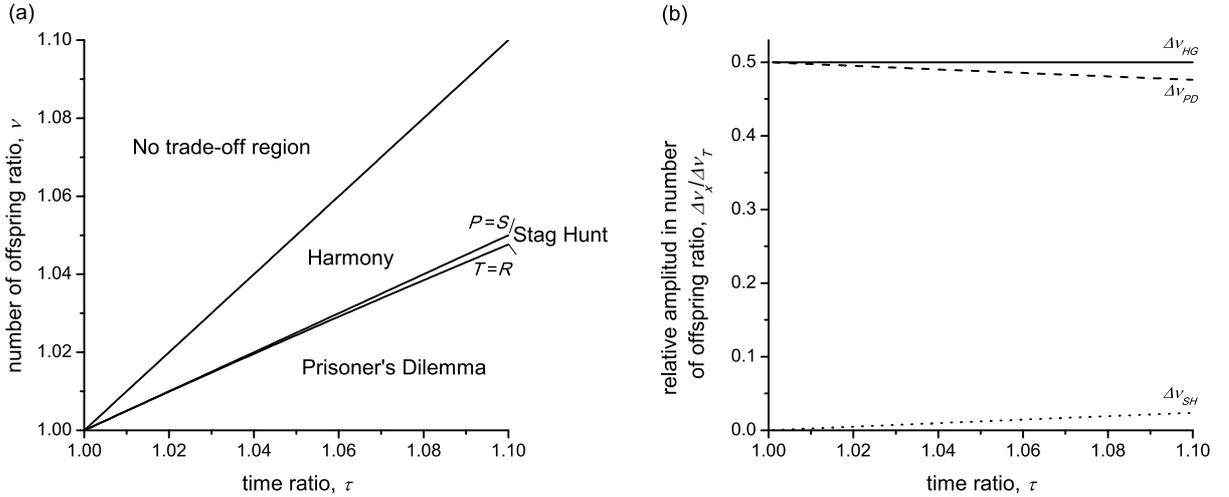}
\caption{Non-cooperative games determined by the rate versus yield trade-offs in the weak selection limit. (a) The three kinds of games determined by rate versus yield trade-offs are shown as a function of the time ratio $\tau$ and the number of offspring ratio $\nu$, where the trade-off condition imposes $\tau>\nu>1$. As it can be observed, in the weak selection limit the Stag Hunt region reduces, and the games occupying most of the parameters space are the Prisoner's Dilemma and the Harmony Game, as for the case of direct interactions between individuals. (b) The relative amplitud in $\Delta \nu$ units occupied by each game is shown as a function of the time ratio $\tau$. As it can be observed, for the weak selection limit shown here, the predominant games are the Prisoner's Dilemma and Harmony Game.  \label{fig:RvsY_2}}
\end{center}
\end{figure}
Whenever the new strategy is only slightly different from the main behaviour, the stag hunt region reduces and practically disappears, as it can be observed in Fig.~\ref{fig:RvsY_2}. Thus, new mutants will be selected for or disappear irrespective of the frequencies at which they appear: if the new mutant strategy is a yield maximizer (cooperator) and determines a harmony game, it will be selected for and invade the population, whereas if it determines a prisoner's dilemma, it will be counter selected and will disappear; if the new mutant strategy is a rate maximizer (defector), the same will happen to it, but being selected for if it determines a PD, counter selected otherwise. Thus, the appearance of the stag Hunt region for the rate versus yield trade-off, which was not present for direct interactions between individuals (see section~\ref{sec:onlyPDandHG}), may be considered negligible in real situations where the weak selection limit may be assumed, being again the prisoner's dilemma and the harmony game the major situations, as for the case of direct interactions. 

\subsection{The competitive exclusion principle}
The competitive exclusion principle, also called Gause's principle, was first discussed by G.F. Gause in 1934 \cite{gause:1934} regarding previous works of Lotka and Volterra, and made famous by Hardin some decades later \cite{hardin:1968}, who stated it in its maxim form: ``complete competitors cannot coexist''. The competitive exclusion principle thus states that, whenever there are two species occupying the same ecological niche, e.g. using the same resources, and living in the same area, even the slightest difference in reproductive rates will lead one of the species to extinction. In game theoretical terms this is equivalent to individuals playing prisoner's dilemmas, harmony games and stag hunt games in well-mixed populations, where coexistence is not possible, and evolution leads to the survival of only one species or behaviour. 

As we have seen in section~\ref{sec:onlyPDandHG}, whenever there are direct interactions between individuals, the principle is true, as only prisoner's dilemmas and harmony games happen in that case, and therefore whenever there are two kinds of behaviours, one of them is led to extinction. It was also shown in section~\ref{sec:coexistence} that the interplay between three different behaviours and mutations allows for coexistence of the three species in higher levels than predicted by the mutation term. Note however that this does not represent an exception to the competitive exclusion principle, as mutations between different behaviours happen, and thus they do not represent different species, but different behavioural types of the same species.

In section~\ref{sec:RvsY} it was shown that trade-offs between rate and yield in resources use determine prisoner's dilemmas, stag hunt games and harmony games in well-mixed situations. In this case again, irrespective of the region in the parameters space where the system lies, the game structure does not allow for stable coexistence between strategies, and thus the competitive exclusion principle works, leading one of the behaviours to extinction. However, some recent studies have shown that a certain type of self-organizing process which roots on a feedback between resources and population composition, may allow for stable coexistence of two strategies in the case of direct interactions in well-mixed populations, and in the absence of mutations \cite{requejo:2012,requejo:2013,requejo:2012c}, being a theoretical example of a situation where the competitive exclusion principle does not work. These results, are reviewed in the next section.

\subsection{An exception to the competitive exclusion principle: An internal self-organizing process allowing for coexistence of cooperators and defectors.}
In words of Hardin \cite{hardin:1960} ``The ``truth'' of
the --competitive exclusion-- principle is and can be established
only by theory, not being subject to
proof or disproof by facts, as ordinarily
understood''. 
With these words, he referred to the fact that, even if an experimental setting is designed and used to prove that during a certain period of time there is coexistence between two species occupying the same ecological niche, and living in the same area, it is not possible to assure that the competitive exclusion principle is not at work, as coexistence may just be a transitive effect, and extinction of one of the species may happen if longer experiments are carried out. And as we do not have infinite time to carry out experiments, the validity of the competitive exclusion principle cannot be disproved experimentally. However, he left a door open to prove it wrong theoretically. In what follows I review theoretical and numerical evidence of what may be regarded as an exception to the competitive exclusion principle.

Recent studies have presented extensions to the mathematical framework of evolutionary game theory in order to study the effect of a limited amount of resources on the evolution of cooperation, using the parasite versus free-rider problem as model of study. In these models, in addition to the frequency dependent selection, a dependence on available resources has been introduced \cite{requejo:2011,requejo:2012,requejo:2013,requejo:2012c}. The models assume that the interactions between individuals are direct and payoffs additive, and that the population is in the mean field limit, i.e. well-mixed. As shown in section~\ref{sec:PD}, this kind of interactions lead only to prisoner's dilemmas and harmony games. However, the models present a situation in which payoffs are non-constant, but depend on available resources. 

In these models there is an influx of resources which is equally distributed between individuals. Such resources are necessary for reproduction, but may also be used in order to carry out parasitic acts on the co-player. The payoff matrix resulting from the interactions between individuals, written using the notation of the present review, is
\begin{equation}
\label{eq:Rmatrix}
\begin{tabular}{l|cc}
 & Y & R \\
\hline
Y & $0$ & $-p b'_p$ \\
R & $p (b'_p - c_p)$ & $-p c_p $
\end{tabular} 
\end{equation}
where the term $1\ge p > 0$ accounts for the constrained capacity of parasites to carry out parasitic actions depending on the amount of resources they possess (note that it cannot be null as there is an influx of resources), and the term $b'_p \le b_p$ accounts for the fact that the receiver of the parasitic action may have lower resources than the maximum a parasite is able to take from the co-player (as before, $c_p$ is the parasitic cost). This matrix may be rewritten without changing the dynamics (using the properties proven in section~\ref{sec:Invariances}) as
\begin{equation}
\label{eq:Rmatrix2}
\begin{tabular}{l|cc}
 & Y & R \\
\hline
Y & $0$ & $a = -p (b'_p-c_p)$ \\
R & $b= p (b'_p - c_p)$ & $0$
\end{tabular} 
\end{equation}
As $p>0$, and thus the matrix cannot become a null matrix due to this term, the situation is analogous to that studied in section~\ref{sec:onlyPDandHG} whenever $b'_p-c_p \ne 0$, and only PD and HG structures happen in this case. If $b'_p-c_p = 0$, then neutral stability may be attained.

\begin{figure}
\begin{center}
\includegraphics[width=160mm]{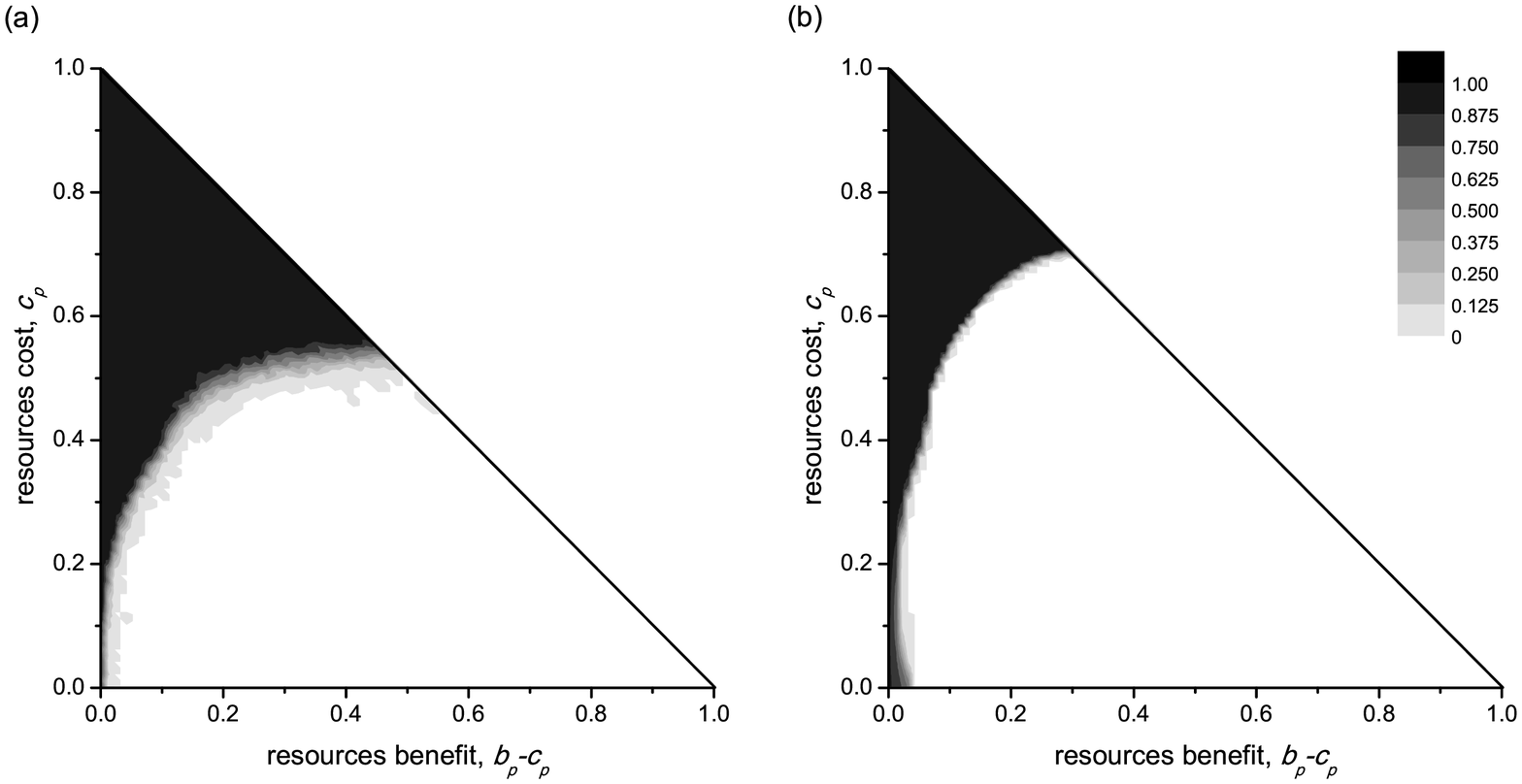}
\caption{Evolutionary outcome of a population of replicating cooperators (free-riders) and defectors (parasites) when resources are necessary for reproduction and survival. The grey scale represents the fraction of cooperative individuals at the end of the evolutionary process as a function of the resources net benefit $b_p-c_p$ and cost $c_p$ of the parasitic action, in units of $r$, the amount of resources necessary for reproduction. The results show two major regions: in white, free-riders extinguish; in black, cooperators extinguish. A dissipation of resources per individual and time step of (a) 0.2$r$ and (b) 0.01$r$ is assumed.   \label{fig:SM}}
\end{center}
\end{figure}
Extensive agent based simulations have been carried out \cite{requejo:2011,requejo:2012} studying two cases: a first one where resources are necessary for reproduction and for survival of the individuals (a certain dissipation of resources is assumed for alive individuals), and a second one where resources are not necessary for the survival of the individuals and deaths occur at random at a rate $f$. The results showed that, whenever resources are necessary for the survival of the individuals \cite{requejo:2011}, two major regions appear, one in which cooperators extinguish, and a second one where defectors extinguish, in accordance to the expected outcome for direct interactions in PD's and HG's regions (Fig.~\ref{fig:SM}). However, things change when resources are only necessary for reproduction.

\begin{figure}
\begin{center}
\includegraphics[width=180mm]{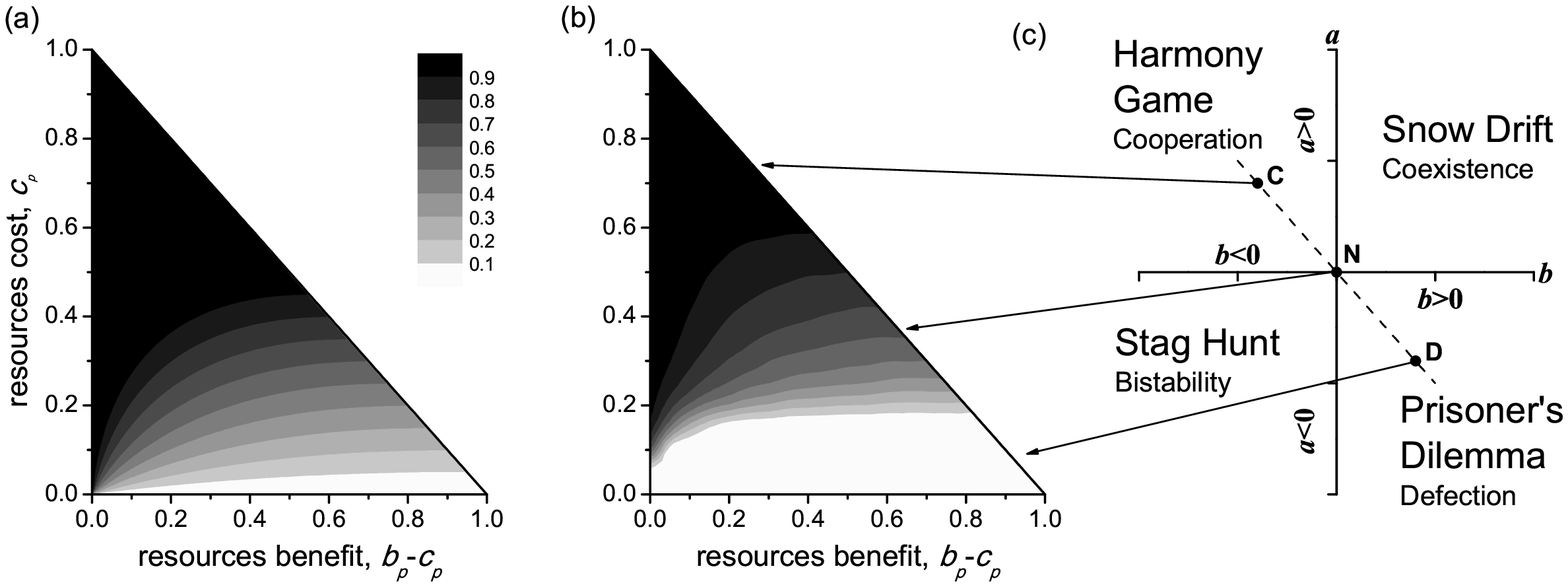}
\caption{Evolutionary outcome of a population of replicating cooperators (free-riders) and defectors (parasites) when resources are necessary for reproduction. The grey scale represents the fraction of cooperative individuals at the end of the evolutionary process as a function of the resources net benefit $b_p-c_p$ and cost $c_p$ of the parasitic action, in units of $r$, the amount of resources necessary for reproduction. The results show the appearance of regions of stable coexistence. (a) Analytical prediction using simplifying assumptions ($f\to0$) as shown in \cite{requejo:2012}. (b) Results of agent based simulations for $f=0.01$. (c) Phase diagram showing the line where the payoff matrix lies, as well as three points: C, corresponding to harmony games, where cooperation is the evolutionary outcome, D, corresponding to a prisoner's dilemma where cooperation extinguishes, and N, corresponding to the neutral matrix. The feedback between resources and population composition allows the system to self-organize for a wide range of parameters (grey regions) and drive the system to point N, i.e. neutral stability.  \label{fig:RD}}
\end{center}
\end{figure}
If deaths occur at random independently of the species (in the analytical treatment it is assumed that the frequency of deaths is negligible compared to that of interactions, i.e. $f \to 0$) and resources are just necessary for reproducing and interacting, then the system undergoes a transition for some parameter values, which allows for stable coexistence of the two species (Fig.~\ref{fig:RD}). This effect, as discussed by Requejo and Camacho \cite{requejo:2012}, roots in a self-organizing process which makes the payoff matrix in Eq.~\ref{eq:Rmatrix2} evolutionary neutral in the coexistence state, which is, moreover, the attractor of the system. In terms of the replicator equation, a third solution appears, triggered by the feedback between resources and population composition. Such feedback may be assumed to root on the dependence of the benefit $b_p$ of the parasitic strategies on the fraction of cooperative individuals in the population. Assuming as a first approximation a linear relationship, $b_p=\alpha \rho$, the replicator equation \cite{requejo:2012} takes the form
\begin{equation}
\label{eq:repeqRD}
\frac{d\rho}{dt} = -\rho (1-\rho )p(b'_p-c_p) = p\rho (1-\rho )(c_p-\alpha \rho).
\end{equation}
which provides a stable coexistence solution whenever $c_p/\alpha \in (0,1)$ (Fig.~\ref{fig:RD}).

\begin{figure}
\begin{center}
\includegraphics[width=160mm]{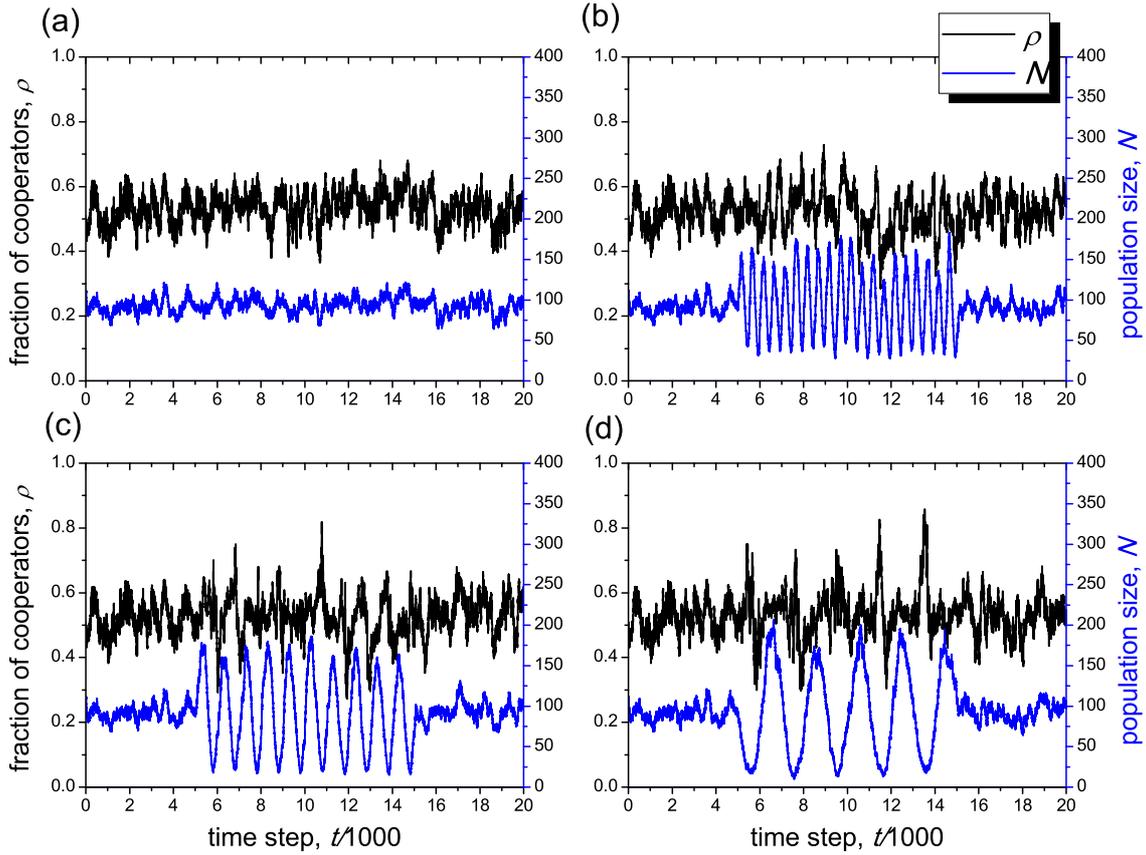}
\caption{Effect of variable influxes of resources in the evolution of a population of cooperators and defectors with a mean population size of $N=93$ individuals. A sinus function $r_{in}=4000+3500 sin(2\pi t/w)$ is assumed to drive the average influx of resources between $t=5000$, and $t=10000$ (some randomness is always allowed), with periods (a) $w=0$, (b) $w=500$, (d) $w=1000$ and (d) $w=2000$. As it can be observed, the population size follows the same behaviour as the variable influx of resources, while the effect on the fraction of cooperators seems to be an increase in noise, and transient promotions of cooperation and defection while the population size decreases and increases, respectively (see Fig.~\ref{fig:RDentre5}). \label{fig:RDsinus}}
\end{center}
\end{figure}
\begin{figure}
\begin{center}
\includegraphics[width=160mm]{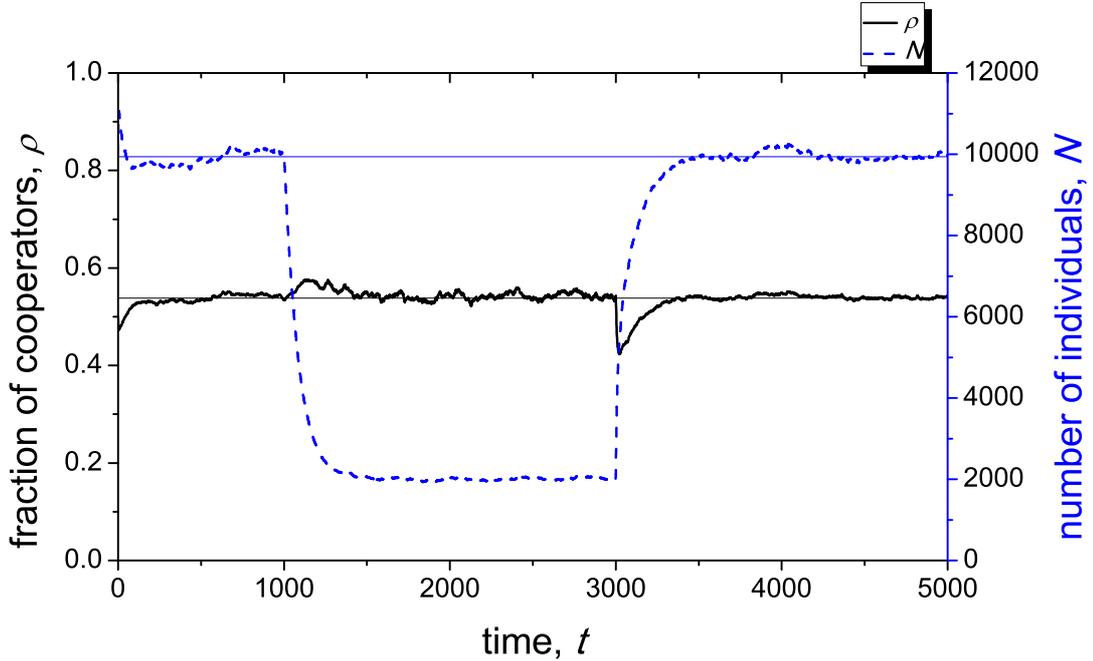}
\caption{Effect of sudden increases and decreases in resource influx on the evolution of a system of cooperators and defectors in a coexistence state. The temporal evolution of such a system is shown, starting in a coexistence state. At $t=5000$ the amount of resources is divided by 5, leading to a short transient in which the fraction of cooperators increases while the population size is decreasing. At $t=10000$ the amount of resources returns to its original value, which leads to a pronounced decrease in cooperation during a short period of time, while the population size is increasing. \label{fig:RDentre5}}
\end{center}
\end{figure}

The stability of the system has also been tested in small populations introducing oscillating amounts of incoming resources. As it can be observed in Fig.~\ref{fig:RDsinus}, the effect of a variable amount of resources does not alter the coexistence state, its influence being only on the population size under study, which is proportional to the amount of resources entering the system. Furthermore, sudden increases or decreases in the amount of resources seem to produce only a transient effect on the coexistence state, being cooperation slightly promoted while the system re-adapts to a sudden decrease in resources, and defection being promoted by sudden increases of resources, as shown in Fig.~\ref{fig:RDentre5}. The last case, when resources suddenly increase, may represent situations in which a new source of resources is found; as it is shown, new sources of resources promote parasitism while the population size increases towards the new stable state.

\subsubsection{Simplified analytical models and phase transitions in evolution}
Simplified analytical models have been developed to describe analytically the effects observed in the previous models \cite{requejo:2012b,requejo:2013}, which relied primarily on agent based simulations, and accounted for a continuum range of resources which individuals possessed. The simplified models make similar assumptions as the previous models, but assume that individuals do not posses an amount of resources belonging to a continuum, and instead may or may not have resources. Furthermore, some new parameters are introduced in order to allow for the analytical treatment. These parameters include --the rest of the parameters are explained below-- $\alpha$, accounting for the fraction of effective interactions between defectors able to carry out parasitic actions and other individuals, and $q$, the probability that a defector looses its resources as a consequence of its behaviour, and which may be interpreted as the mean cost per interaction $c_p$ of the previous model. This allows to write a set of differential equations which may be solved numerically \cite{requejo:2012b,requejo:2013}, and which show the phase transitions that occur in the system. The models account for three situations: one where resources are necessary for reproduction and survival, one where they are only necessary for survival and deaths occur at random, and a third one where resources are necessary for survival, deaths occur at random, and the population size is constant.

\begin{figure}
\begin{center}
\includegraphics[width=160mm]{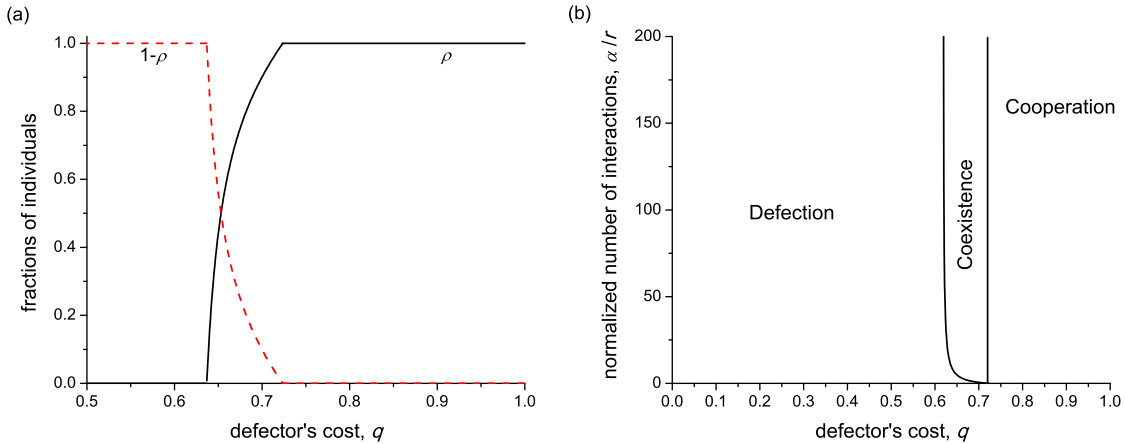}
\caption{Phase transitions when resources are necessary for survival. (a) Fractions of cooperators $\rho$, (solid line) and defectors $1-\rho$ (dashed
line) as a function of parameter $q=c_p$ for $\alpha/r$ = 10, where $\alpha$ is the probability of acting of defectors in active states and $r$ the resources dissipation rate of individuals to keep alive. Two phase transitions happen, allowing for three different regions: dominance of defection, coexistence and dominance of cooperation. (b) Phase diagram showing the regions of defection and coexistence as a function of $q$ and $\alpha/f$. \label{fig:SMsimp}}
\end{center}
\end{figure}
The analysis of such models shows that whenever resources are necessary for reproduction and survival, two phase transitions happen, the first one from a defective state to a coexistence state between cooperators and defectors, the second from the coexistence state to dominance of cooperation. It has been proven \cite{requejo:2012b} that the population composition can be written in the simple form
\begin{equation}
\label{dim}
C_i,D_i=g_i(q,\frac{\alpha}{r})\frac{E_T}{r},
\end{equation}
where $C_i$ and $D_i$ are the fractions of cooperators and defectors with $i$ resources, $i=0,1$, $g_i$ are functions of the corresponding parameters, and $E_T/f$ is the quotient between total incoming resources $E_T$ and death rate $f$, which sets the population size. Thus, the transitions towards cooperation depend on the parameters $\alpha$ and $q$ presented in the previous paragraph, as well as on $r$, the rate of dissipation of resources of alive individuals. As it can be observed in Fig.~\ref{fig:SMsimp}, the region of coexistence is small compared to the regions of dominance of each strategy, being cooperation dominant for high parasitic costs, $q>0.72$, as calculated in \cite{requejo:2012b}. 

\begin{figure}
\begin{center}
\includegraphics[width=160mm]{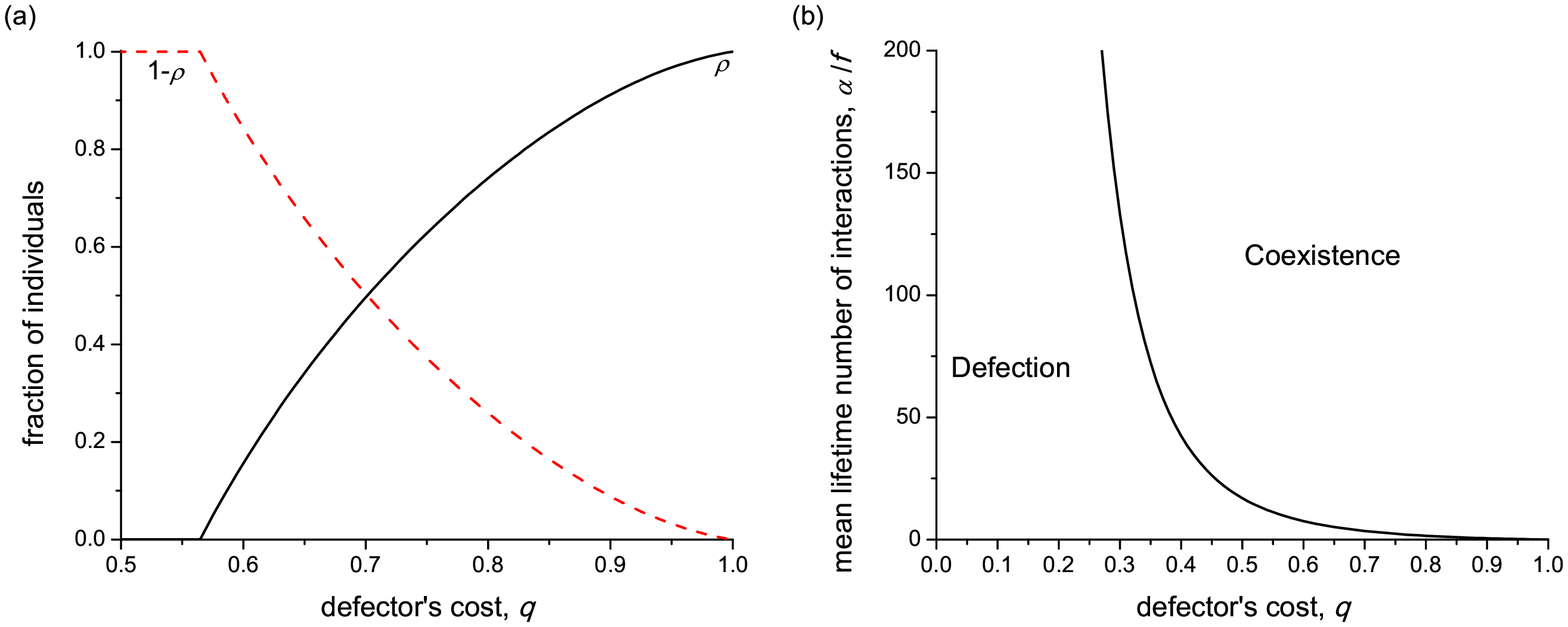}
\caption{Phase transition to coexistence between cooperators and defectors when resources are not necessary for survival. (a) Fractions of cooperators $\rho$, (solid line) and defectors $1-\rho$ (dashed line) as a function of parameter $q=c_p$ for $\alpha/f$ = 10, where $\alpha$ is the action probability of defectors in active states and $f$ the death rate. Below a critical value, cooperators extinguish. (b) Phase diagram showing the regions of defection and coexistence as a function of $q$ and $\alpha/r$. \label{fig:RDsimp}}
\end{center}
\end{figure}
When resources are necessary only for reproduction and deaths occur at random at a rate $f$, a phase transition happens from a fully defective population to coexistence states, as shown in Fig.~\ref{fig:RDsimp}. In this case the population composition may be written as
\begin{equation}
\label{dim2}
C_i,D_i=h_i(q,\frac{\alpha}{f})\frac{E_T}{f},
\end{equation}
The transition depends on the parameters $\alpha$ and $q$, as well as on $f$, the death probability. As it can be observed in Fig.~\ref{fig:SMsimp}, the region of coexistence increases with increasing the quotient $\alpha/f$, i.e. with increasing the mean lifetime number of interactions.

\begin{figure}
\begin{center}
\includegraphics[width=160mm]{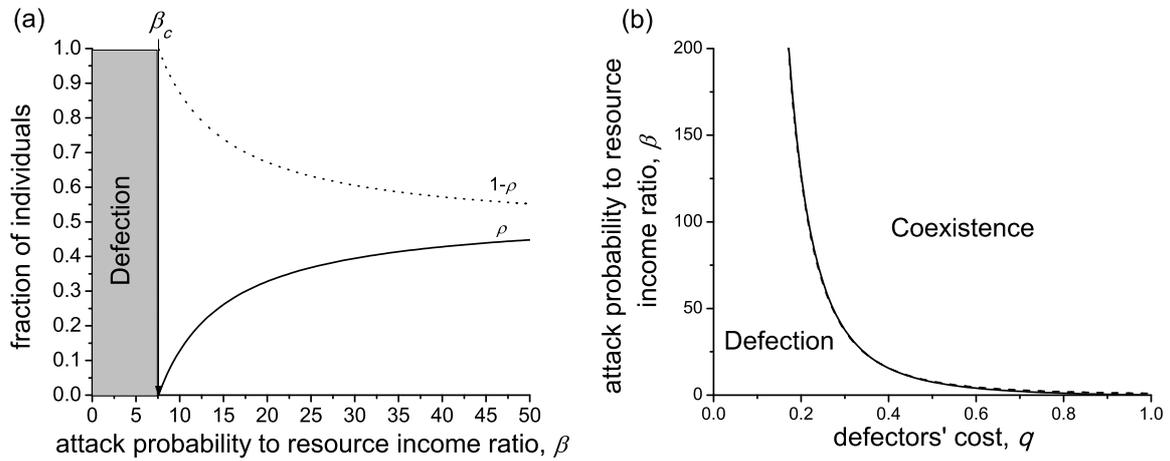}
\caption{Phase transition to coexistence between cooperators and defectors when resources are not necessary for survival and population size is constant. (a) Fractions of cooperators $\rho$, (solid line) and defectors $1-\rho$ (dashed line) as a function of parameter $c_p$ for $\beta=\alpha/\gamma$ = 10, where $\alpha$ is the action probability of defectors in active states and $\gamma$ the resource income rate. Below a critical value, cooperators extinguish. (b) Phase diagram showing the regions of defection and coexistence as a function of $q$ and $\beta$. \label{fig:RDsimpNcte}}
\end{center}
\end{figure}
When resources are necessary only for reproduction and deaths occur at random, but at a rate which maintains population size constant, a phase transition happens from a fully defective population to coexistence states, as shown in Fig.~\ref{fig:RDsimpNcte}. This transition depends on parameters $\beta=\alpha/\gamma$ and $q$, being $\gamma$ the resource income rate. Above the threshold parameters at which the transition happens, the population composition can be written as
\begin{equation}
\label{eq:mixed}
C_i= a_i (1-\frac{\beta_c}{\beta}),  \hspace{1cm} D_1=\frac{a_2}{\beta},
\end{equation}
with $i=0,1$, and $a_i$ and $\beta_c$ functions of parameter $q$ (note that, being the population size constant, $D_0$ may be computed from these quantities). 
Remarkably, in contrast with the situations in which the population size varies and an increase in resources just changes the population size, but not the final outcome of the system, in this case in which the population size is constant a decrease in resources income rate $\gamma$ is enough to trigger the transition to coexistence states between cooperators and defectors (see Fig.~\ref{fig:RDsimpNcte}.

\section{Conclusions}
\label{sec:discussion}

The evolutionary game theoretical framework has expanded greatly during the last decades, and the study of the evolution of cooperation, a paradigmatic issue since the very beginning of the evolutionary thinking, has reached contexts which where considered as science-fiction when it all began. The study of the simple rules that allow for the promotion of cooperation has been superseded by new studies that include some of the intricacies of an ecological and economical world in which feedbacks, synchronizations and self-organizing processes give rise to unexpected behaviours. However, as it has been shown in the present review, taking care of the simple concepts that triggered the development of the framework is still a source of surprising results, as the fact that random exploration of strategies determining prisoner's dilemmas allows for the survival of all of them, or that the limitation of resources may trigger coexistence states in a situation in which following the competitive exclusion principle one would say that it should not happen. For this reason, and given the expanding complexity of the models, which include more and more details of the great reality surrounding us, the present review went back to the beginning of the path: first, to the history, the relevant concepts and their evolution, and then to the roots of the problem, the struggle for life, and inherently to it, the struggle for resources. This last point seems specially important in a world in which sources of resources which where assumed to be infinite during the industrial revolution are at risk of becoming exhausted, and in which the atmosphere, the seas and all ecosystems, are starting to change due to human behaviour. Although much more accurate models are necessary to represent such complex situations, the present review presented some results that, though very simple, may allow for the development of such models, and ultimately for a better understanding of the world we live in and the society we are creating. 

\section{Acknowledgements}

I specially thank Juan Camacho for his patience and wise advise during the development of my Ph.D. thesis, when part of this work was carried out, as well as for the fruitful collaboration that gave rise to part of the work that I review here. I also thank Hanna Kokko for useful comments and discussion on the first manuscript of the paper, and Jos\'e Cuesta and Alex Arenas for interesting conversations during which some of the ideas included in this review came into existence. I acknowledge the logistic support provided by the Statistical Physics group at the Universitat Aut\`onoma de Barcelona and the Evolution, Ecology and Genetics group at the Research School of Biology (the Australian National University). Financial support was provided by the Ministerio de Ciencia y Tecnolog\'{i}a (Spain) (FPU grant; project FIS2009-13370-C02-01) and the Generalitat de Catalunya (project 2009SGR0164).

\begin{appendix}

\section{Reference strategies, dynamical equality and distinguishablity}
\label{sec:DynamicalEquality}
If the evolutionary rule depends on the payoff difference between strategies, as it happens for some local rules \cite{szabo:1998, traulsen:2006b}, or between the strategy's payoff and the mean population fitness, as for the replicator equation \cite{hofbauer:2003}, the dynamics is invariant under the addition of a constant to all payoffs (see Sec.\ref{sec:Invariances}). Suppose now that there are two different situations, one represented by payoff matrix $\Pi$, corresponding to equation~\eqref{eq:PDmatrix}, the other one by $\Pi{}'$. We may add a constant $k$ to the second one, and find the necessary conditions to have equal dynamics, i.e. find the parameters that fulfil 
\begin{equation}
\label{eq:equality}
\Pi = \Pi{}' + k.
\end{equation}
By doing so, one finds
\begin{equation}
\label{eq:conds}
\begin{tabular}{l}
$\Delta_s = \Delta_s'$ \\
$\Delta_r = \Delta_r'$ \\
$k = A_r - A_r' + A_s - A_s'$
\end{tabular}
\end{equation}
This conditions tell us when the dynamics of the two situations are indistinguishable given the previous assumptions, and thus cannot be used to tell the difference between both scenarios. Note that if we are dealing with direct interactions between individuals, which produce a fitness change in actor and recipient, and fitness is additive, then equations~\eqref{eq:Cgames},~\eqref{eq:PD} must hold, which is consistent with equations~\eqref{eq:conds}.

Now, suppose that we are dealing with the more interesting case in which one of the strategies is present in both situations, and might be used as reference to establish a relative scale of cooperativeness. If this strategy is the cooperative one in both scenarios, then the only possibility that fulfils equations~\eqref{eq:conds} is that the defective strategy is also the same, and $k=0$. Both situations are then not only indistinguishable by looking to the dynamics, but the same indeed. 

There is a more interesting case, however, when the strategy present in both scenarios is regarded as defective in one case, as cooperative in the other, i.e. $B_s = A_s', B_r = A_r'$. In this case, this strategy may be used as reference. Equations~\eqref{eq:conds} reduce to
\begin{equation}
\label{eq:conds2}
\begin{tabular}{l}
$2B_s = A_s + B_s'$\\
$2B_r = A_r + B_r'$\\
$k = \Delta_s + \Delta_r > 0$
\end{tabular}
\end{equation}
The fact that $k > 0$ comes from the restrictions introduced in equation~\eqref{eq:PD}, both for PD and HG. This has an important implication: Even if the dynamics of two situations are identical (given the assumptions above), we may always measure a baseline fitness difference $k$ to tell them apart. Furthermore, the equations tell us the relationship between cooperativeness and selfishness degrees ($\Delta_r, \Delta_s$), and difference in mean population fitness $k$. 

It might not seem really surprising the fact that we can always find a difference when the systems are not equal (even if the dynamics are indistinguishable). However, the fact that we may quantify such difference measuring differences in baseline fitness, and relate it to higher or lower cooperativeness and selfishness of the interactions between the different behaviours or species, allows us to define a relative scale for cooperation.

	\section{More on the definitions of cooperation and altruism}
	\label{sec:definitions}
\subsection{Definitions of cooperation and altruism along the literature}
	Some of the definitions in the literautre are summarised here.
	\begin{itemize}
	\item ``\emph{The degree of co-operation observed in nature varies along a continuum, from the one extreme of severe parasitism/virulence to the other extreme of mutual benevolence. [\ldots]  An observed level of co-operation
requires evolutionary explanation only in so far as that level deviates from a level representing the "null" point for the species interaction, and calculation of this null point is somewhat subjective. [\ldots] Our designation of a phenotype as cooperative need only imply that it is more co-operative than some feasible alternative.}'' Bull and Rice, 1991 \cite{bull:1991} \\
This definition refers to relative cooperation.

	\item ``\emph{Cooperation is an outcome that --despite potential relative costs to the individual-- is "good" in some appropriate sense for the members of a group, and whose achievement requires collection action. But the phrase "to cooperate" can be confusing, as it has two common usages. To cooperate can mean either: (1) to achieve cooperation--something the group does, or (2) to behave cooperatively, that is, to behave in such a manner that renders the cooperation possible (something the individual does), even though the cooperation may not actually be realised unless other group members also behave cooperatively.}'' Dugatkin 1997 \cite{dugatkin:1997} \\
The difference between cooperative behaviour and cooperation is clear in this definition.

	\item ``\emph{The key distinction we wish to make is between cooperation (an interaction between two or more individuals) and cooperative behaviour
(an action or actions taken by a single individual). [\ldots] We define cooperation as an interaction between individuals that results in net benefits for all of the individuals involved}'' Bergmueller et al. 2007 \cite{bergmuller:2007} \\ This definition is slightly more restrictive than the one in the main text, though both overlap if one accounts for peace as a social good or benefit.

	\item Referred to lifetime consequences: ``\emph{Cooperation: A behaviour which provides a benefit to another individual (recipient), and which is selected for because of its beneficial effect on the recipient}'' West et al. 2007 \cite{west:2007} \\ 
This definition mixes cooperation, which is something carried out by at least two interacting individuals (see the definition in reference \cite{dugatkin:1997} above), and cooperative behaviour, i.e. individual behaviour which allows for cooperation.

	\item ``\emph{- Cooperative behaviour: a behaviour that on average increases the fitness of a recipient and which is under positive selection if it on average increases the inclusive fitness of the actor via direct fitness benefits. [\ldots] - Altruistic behaviour: a behaviour that on average
increases the fitness of a recipient and which is
under positive selection if it on average increases
the inclusive fitness of the actor via indirect fitness
benefits. [\ldots] - Cooperation: two (n) partners increase on average
their direct fitness due to the interaction.}'' Brosnan and Bshary 2010 \cite{brosnan:2010} \\
The first definition above accounts always for cooperative behaviours (behaviours that, when interacting together, create direct fitness benefits for all interacting individuals). However, the above definitions are blurry in some situations, as when behaviours are counter-selected, e.g. if an actor provides a benefit to a recipient, and the inclusive fitness of the actor increases due to direct fitness benefits, but less than the average increase in the population, it would be under negative selection, and the above definition cannot be applied to call it cooperative or not. Furthermore, according to the definitions altruistic behaviours are not considered cooperative, nor cooperative behaviours altruistic.
	\end{itemize}

\subsection{Remarks on the definitions in the text}
\label{app:remarks}
The present review starts with the statement of a relative definition of cooperation and selfishness, avoiding commentaries of whether cooperation is intentional or unintentional. The definition of cooperation in this way includes by-product mutualism, situation which might be regarded as a case of unintentional cooperation. Some authors have claimed that this is on the limit of the scope of cooperation \cite{bull:1991,west:2007}, and that it is important to understand the emergence of stable ecosystems and other selective units \cite{maynard-smith:1995a} from cooperative interactions, independently of their intentionality. 

As an example, an elephant excreting dung is acting beneficially for itself, and for beetles feeding on such dung. Some authors argued that such behaviour is out of the scope of cooperation, as the situation is a one-way by-product benefit, and the elephant behaviour does not deviate from the behaviour found in the absence of beetles, which in this case represents the null point \cite{bull:1991,west:2007,brosnan:2010}. However, it is known that baby elephants eat other elephants dung in order to obtain some bacteria that feed on it, and which they need to incorporate to their intestine in order to digest the vegetation present in the savannah and jungles. According to the same reasoning, if an increase in dung leads to an increase of such bacteria, which in turn allows elephants to produce more dung, this second situation might be classified as cooperative.

The provided definition explicitly states that cooperation must not be forced. Other authors require it to be voluntary to prevent some exploitative or slaver behaviours and acts to be regarded as cooperative, as food in exchange for work or forced starvation otherwise, which might be found in the sometimes wrongly-classified as mutualistic interactions between ants and aphids, where ants take care of aphids as far as they provide them with food, but kill them otherwise. Requiring cooperation not to be forced allows for coherence with this argument, as well as with those in the previous paragraphs for including non-intentional behaviours.

Cooperative acts may also be carried out in big groups \cite{dugatkin:1997}, where one action has many simultaneous recipients. According to the definition, we might talk about cooperative acts of one individual directed to another, as it might happen that the action of an individual has non-negative effects and negative effects on different individuals at the same time; e.g. suppose the elephant dung falls on an ants nest and blocks the entrance: The act would be cooperative for the beetles and bacteria using the dung, while non-cooperative for the ants.

\end{appendix}

\end{document}